\documentclass[a4paper,11pt]{article}
\pdfoutput=1 

\usepackage{jheppub} 

\usepackage[T1]{fontenc} 
\usepackage{makecell}
\usepackage{appendix}
\usepackage{amsthm}
\usepackage[all,cmtip]{xy}
\usepackage{amsfonts,amssymb,epsfig,amsmath,mathtools}
\usepackage[utf8]{inputenc}
\usepackage{textcomp,setspace}
\usepackage{tabularx,bm,booktabs,multirow,tabu}
\usepackage[inline]{enumitem}
\usepackage{hyperref,caption,subcaption}
\usepackage{xcolor,tikz,graphicx}
\usetikzlibrary{shapes.geometric,positioning}
\usetikzlibrary{cd}
\usepackage{verbatim}

\newtheorem{mydef}{Definition}
 
\newtheorem*{conj}{Conjecture}
\newtheorem*{thm}{Criterion}

\title{ \bf Bootstrapping ADE M-strings}

\author{Zhihao Duan}
\author{and June Nahmgoong}

\affiliation{School of Physics, Korea Institute for Advanced Study, Seoul 02455, Korea}

\emailAdd{xduanz@gmail.com}
\emailAdd{junenahmgoong@gmail.com}

\abstract{ We study elliptic genera of ADE-type M-strings in 6d (2,0) SCFTs from their modularity and explore the relation to topological string partition functions. We find a novel kinematical constraint that elliptic genera should follow, which determines elliptic genera at low base degrees and helps us to conjecture a vanishing bound for the refined Gopakumar-Vafa invariants of related geometries. Using this, we can bootstrap the elliptic genera to arbitrary base degree, including D/E-type theories for which explicit formulas are only partially known. We utilize our results to obtain the 6d Cardy formulas and the superconformal indices for (2,0) theories.}

\preprint{KIAS-P20045}

\begin{document} 
\maketitle
\flushbottom

\section{Introduction}

Recently we have witnessed a revival of interest in six-dimensional quantum field theories. In particular, 
6d $\mathcal{N} = (2,0)$ superconformal field theories (SCFTs) play an important role in our understanding of lower dimensional dualities \cite{Gaiotto_2012,Alday:2009aq,Dimofte_2013,Gadde_2016}. Although still lacking Lagrangian descriptions, they are well-known to be labeled by simply-laced, i.e., ADE Lie algebras, and allow for self-dual string solitons.

Historically, 6d (2,0) SCFTs were first constructed from type IIB superstring compactifications \cite{Witten:1995zh, Seiberg:1996vs}. In this paper, we consider a more unifying approach, which involves F-theory \cite{Vafa_1996,Morrison_1996,Morrison_1996II}. Roughly speaking, F-theory is a non-perturbative completion of type IIB string theory, by allowing the non-constant axion-dilaton field over the moduli. Its VEV can be dictated by the complex structure of an elliptic curve, which hints at the class of Calabi-Yau (CY) manifolds having elliptic fibrations. F-theory compactification on elliptically fibered CY threefolds can engineer six-dimensional supersymmetric gauge theory. In \cite{Heckman_2014, Heckman:2015bfa}, it is conjectured that all possible six-dimensional $\mathcal{N} = (1,0)$ SCFTs can be obtained from this approach. As particular examples, $(2,0)$ SCFTs correspond to CYs with trivial elliptic fibrations.

On the other hand, CY threefolds are central objects in another area known as topological string theory. Given a CY threefold $X$, we first consider a sigma model of a two-dimensional worldsheet with fixed metric mapping to $X$. Its action needs to be topologically twisted in order to be topological. Then we couple the system to two-dimensional gravity, i.e., integrate over all possible metrics and topologies and obtain what is known as the topological string.

Furthermore, people discovered that for local CY threefolds, M-theory compactification engineers 5d gauge theory with $\mathcal{N} = 1$ supersymmetry, as an uplift of the relation between type IIA superstring and 4d physics \cite{Katz:1996fh}. After a sequence of seminal works \cite{Nekrasov:2002qd,Nekrasov,Hollowood:2003cv,IKV}, \cite{Aganagic:2011mi, Huang:2011qx} proposed a refinement of topological string theory, which interprets gauge theory results in terms of geometric invariants known as refined Gopakumar-Vafa (GV) invariants.  For elliptically fibered CY threefolds that engineer 5d theories having UV completions to 6d, the relevant gauge-theoretic quantity is precisely the elliptic genus of self-dual strings. Therefore, understanding it becomes an important task for both communities.

Currently, a plethora of ways to compute the elliptic genera for various 6d SCFTs has already been introduced. Based on their starting point, we roughly divide them into two classes and make a very partial list. From the point of view of higher dimensional gauge theories, one can utilize the brane system if it exists. For example, concerning the E-strings or M-strings, where one considers M2-branes in a chain of M5-branes with or without M9-brane, the domain wall operator method \cite{Haghighat:2013,Haghighat:2014pva,Cai:2014vka} can be used to compute their elliptic genera. Also, if one works out the 2d quiver gauge theories living on the self-dual strings, one can apply localization technique through the Jeffrey-Kirwan residue prescription \cite{Benini_2013,Benini:2013xpa}. In \cite{Haghighat:2013,Haghighat:2013tka,HKLV:2014,Kim:2014dza,Gadde:2015tra,Kim:2015fxa,Kim:2016foj,Kim:2018gjo}, this was successfully carried out to tackle E-strings, M-strings, some minimal SCFTs \cite{HKLV:2014} and many more. From the point of view of topological strings, traditional methods such as the topological vertex or mirror symmetry are not that straightforward to apply. The non-toric topological vertex \cite{Kim:2015jba,Haghighat:2013tka,Hayashi:2017jze,Foda_2018} is developed to compute elliptic genera of E-strings, M-strings, and certain limit of elliptic genera for several other (1,0) minimal SCFTs. B-model technique is explored in \cite{Huang:2013yta,Haghighat:2014vxa}, but restricted to the first few genus expansions. Recently, \cite{Gu_2019,Gu_2019II,gu2019elliptic,Gu:2020fem} found a powerful way to compute the elliptic genera using 6d blow-up equations, which in particular covers all the theories in the non-higgsable clusters \cite{Morrison:2012np}.

Last but not least, since elliptic genus is known to transform as a Jacobi modular form, one could fully explore its modular property. In other words, we write down a modular ansatz with a finite number of undetermined coefficients, discussed in detail in section \ref{sec:bootstrap}, then impose sufficient constraints either from gauge theory or topological string theory to fix the unknowns. This approach is often dubbed modular bootstrap \cite{DelZotto:2016pvm,Gu:2017ccq,DelZotto:2017mee, Kim:2018gak,DelZotto:2018tcj,Duan:2018sqe}, and was successfully applied to compute the elliptic genera of E-strings, M-strings, various minimial SCFTs, conformal matters, etc., at low base degrees, sometimes in the unflavored limit.

However, even though much progress has been made in understanding the $\mathcal{N}$ = (1,0) minimal SCFTs, higher rank (2,0) SCFTs have not yet been fully explored in the current literature. The A-type theory has a clear brane picture as a stack of M5-branes \cite{Haghighat:2013}. However, elliptic genera of strings in D/E-type theories were partially studied in the massless limit only \cite{Gadde:2015tra, Haghighat:2017vch}. In this paper, along the line of the modular bootstrap approach, we initiate a uniform investigation of the elliptic genera of all ADE-type (2,0) SCFTs with all chemical potentials fully refined. Our approach captures the elliptic genera of massive M-strings for D- and E-type, which, as far as we know, have not been written down before.\footnote{We notice that some elliptic genera of D4-type theories were computed in \cite{Gu:2017ccq, Kim:2018gak}.}

We will discuss two different ways to bootstrap the elliptic genera. The first one is based on the fact that after a circle compactification, 6d (2,0) \(\mathfrak{g}\)-type SCFT becomes 5d maximally supersymmetric Yang-Mills (SYM) theory with gauge group $G$. Then, the 6d partition function has an expansion in terms of instanton numbers. Although the instanton partition function is unknown if \(G\) is exceptional, the perturbative part is well-known. Also, from the definition of the elliptic genus, there is a kinematical constraint that we called "flip symmetry" in subsection \ref{sec:flip}. Combining the two, we are able to fix many elliptic genera at low base degrees.

The second approach is based on the property of refined GV invariants. They depend on the curve classes $\alpha \in H_2(X,\mathbb{Z})$ and two spins related to the representation of 5d little group $SO(4)$, so we label them by $N^{\alpha}_{j_-j_+}$. As will be explained in subsection \ref{section4}, there exists a vanishing bound for those invariants. Namely, for a fixed $\alpha$, $N^{\alpha}_{j_-j_+}$ vanishes when either $j_-$ or $j_+$ is sufficiently large. Nevertheless, it is, in general, not easy to derive the precise bound from the geometry.  In subsection \ref{section4}, based on the data obtained from the elliptic genera of low base degrees, we experimentally conjecture a vanishing bound and propose that it should be able to determine the elliptic genera of arbitrary base degrees in principle.

This article is organized as follows. In section \ref{section2}, we give a brief review of ADE-type (2,0) SCFTs, elliptic genera, topological strings and their close relationship. In section \ref{sec:bootstrap}, we discuss two ways to compute the elliptic genera. More specifically, in subsection \ref{sec:flip}, we first introduce the modular ansatz then show how to combine the 5d perturbative part with flip symmetry to bootstrap the elliptic genera. In subsection \ref{section4}, we first argue the existence of vanishing condition and conjecture an empirical bound for GV invariants of related geometries. We also show how to make use of this constraint to bootstrap the elliptic genera recursively. In section \ref{section5}, we give two applications of our results: the 6d Cardy formulas for all (2,0) SCFTs and the (2,0) superconformal index in the \(\mathcal{W}\)-algebra limit. In section \ref{sec:Conclusion}, we finish the paper with concluding remarks.

\section{6d \((2,0)\) SCFTs and geometries}
\label{section2}
In this section, we briefly introduce 6d ADE (2,0) SCFTs and the role of the self-dual strings from the point of view of F-theory. We will further discuss the supersymmetric partition function of the 6d SCFTs on \(\mathbb{R}^4\times T^2\) in the $\Omega$-background and the elliptic genus of the M-strings. Finally, we explain the connection of the whole story to topological strings.

To start with, we discuss the geometric construction of six-dimensional SCFTs. This is achieved through compactifying F-theory on an elliptically fibered CY threefold. In other words, we assume the CY threefold $X$ enjoys a torus fibration $E$ over a non-compact complex surface $B$,
\begin{equation*}
\xymatrix{ T^2 = E \ar[r]&X\ar[d]^\pi\\ 
	&B\\}
\end{equation*}
How does geometry see whether a given field theory can reach a CFT fixed point or not? Remember that although lacking Lagrangian descriptions, six-dimensional quantum field theories have stringy solitons. This can be naturally explained in the geometric setup since a D3-brane wrapped on a two-cycle in $B$ becomes a non-critical BPS string in 6d, whose tension is proportional to the volume of the two-cycle. Such a string is called the `self-dual string' because it inherits the self-duality condition from the D3-brane in ten dimensions. If we assume that all the compact divisors inside $B$ are contractible, then the self-dual string becomes tensionless by shrinking down all such curves, which signals a superconformal field theory.

\begin{figure}[ht!]
    \centering
    \includegraphics{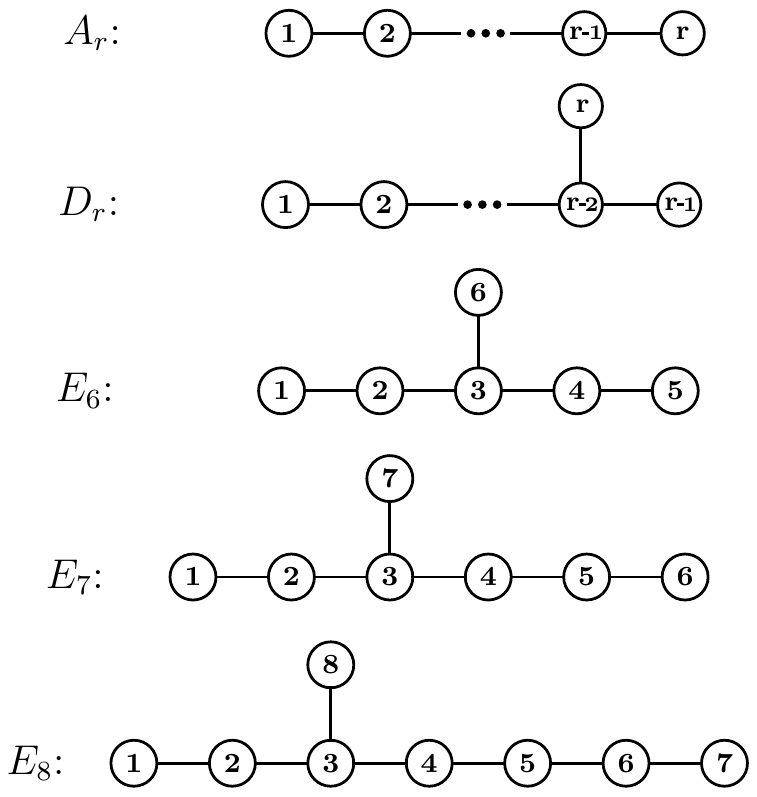}
    \caption{ADE Dynkin diagrams with numbers specifying the order.}
    \label{fig:ADE}
\end{figure}

For cases of our interest, i.e., $\mathcal{N} = (2,0)$ ADE-type SCFTs, the base $B$ in the singular or conformal limit is $\mathbb{C}^2/\Gamma$, with $\Gamma$ the discrete subgroups of $SU(2)$. The famous Mckay correspondence states that there is a one-to-one map between those subgroups and simply laced, i.e., ADE-type Lie algebras. This can be understood as follows. In the orbifold $\mathbb{C}^2/\Gamma$, the origin is singular, which can be resolved by a series of blow-ups. This yields exceptional curves intersecting each other according to the negative of the Cartan matrices of ADE Lie algebras. Also, notice that the resolved spaces, also known as ALE spaces, are themselves CY manifolds, so $E$ is trivially fibered over them. However, $X$ must be deformed to incorporate the M-string mass defined below, making $E$ no longer simply a trivial fibration.  This means that the elliptic fiber also carries non-trivial information, which is an important reason why we choose a more general approach. We present the Dynkin diagrams of ADE Lie algebras in Figure \ref{fig:ADE}, where the numbers inside the node fix our convention on the ordering.

\par 

Next, bearing the above geometric picture in mind, let us put our 6d (2,0) SCFTs on \(\mathbb{R}^4\times T^2\) and define the BPS index. We will declare one circle of \(T^2\) as a temporal circle, and turn on the $\Omega$-background on \(\mathbb{R}^4\) so that the self-dual string is wrapped on \(T^2\) and localized at the origin of \(\mathbb{R}^4\). Then, one can define the supersymmetric partition function on \(\mathbb{R}^4\times T^2\) as follows,
\begin{align}
Z_{\mathbb{R}^4\times T^2}(\tau,\epsilon_{1,2},m,\mathbf{v})=\text{Tr}\Bigr[(-1)^F q^P e^{-\mathfrak{n}\cdot \mathbf{v}} e^{-\epsilon_1 J_1-\epsilon_2 J_2} e^{-(\epsilon_+ +m) R_1-(\epsilon_+ -m) R_2} \Bigr].
\label{eq83_R4T2index}
\end{align}
Here, \(J_{1,2}\) are the two angular momenta on \(\mathbb{R}^4\) which rotate two orthogonal planes in \(\mathbb{R}^4\). \(R_{1,2}\) are the two Cartan charges of \(SO(5)\) R-symmetry of (2,0) supersymmetry. We introduce the corresponding chemical potentials as \(\epsilon_{1,2}\) for \(J_{1,2}\) and \(\epsilon_+ \pm m\) for \(R_{1,2}\), with $m$ a mass deformation henceforth referred as the M-string mass and  \(\epsilon_\pm \equiv {\epsilon_1 \pm \epsilon_2\over 2}\). Also, \(P\) is a KK momentum along the spatial circle in \(T^2\) with the conjugate fugacity \(q\equiv e^{2\pi i \tau}\). Here, \(\tau=i{r_0\over r_1}\) where \(r_0\) is the radius of the temporal circle, and \(r_1\) is the radius of the spatial circle in \(T^2\). Lastly, \(\mathfrak{n}=(n_1,n_2,...,n_r)\) denotes the number of the self-dual strings wrapping on each two-cycle in $B$. For this reason, we also call $\mathfrak{n}$ as a base degree. The conjugate chemical potential is given by \(\mathbf{v}=(v_1,v_2,...,v_r)\) which are proportional to the volumes of the two-cycles.
 
Let us take a closer look at the index. Suppose we consider a 6d (2,0) \(\mathfrak{g}\)-type SCFT, then the following expansion for the \(\mathbb{R}^4\times T^2\) index (\ref{eq83_R4T2index}) exists,
\begin{align}
Z_{\mathbb{R}^4\times T^2}= Z_0 \cdot Z_{T^2} = Z_0 \cdot (1 + \sum_{\mathfrak{n} \neq 0\, \in \mathbb{Z}_{\geq 0}^r} Z_{\mathfrak{n}} e^{-\mathfrak{n}\cdot \mathbf{v}})\,,
\label{eq130_EGEXP}
\end{align}
which is valid when the string tension \(\mathbf{v}\) is large. The prefactor \(Z_0\) comes from the KK tower of BPS particles decoupled from the self-dual strings. Its form will be given in section \ref{sec:bootstrap}. On the other hand, the expansion coefficient \(Z_{\mathfrak{n}}\) denotes the contribution from the \(\mathfrak{n}\) number of self-dual strings on \(T^2\). Therefore, it is given by the elliptic genus of the 2d (4,4) SCFT on the \(\mathfrak{n}\) self-dual strings \cite{Haghighat:2013}.
\par 
The elliptic genus is well known to be a weight zero Jacobi modular form (definitions can be found in appendix \ref{appendixJMF}). Thus what remains to determine is its index. It turns out that the index is tightly related to the anomaly polynomial of BPS strings, which we explain below.

To start with, the 6d $\mathcal{N}$ = (2,0) theory has various 't Hooft anomalies, summarized in terms of the eight-form anomaly polynomial. Then based on the anomaly inflow mechanism, \cite{Kim:2016foj,Shimizu:2016lbw} successfully computed the general four-form anomaly polynomial on the 2d worldsheet of BPS strings. In our case, it gets simplified,
\begin{align}
I_4=\sum_{a,b=1}^{r_{\mathfrak{g}}}{\Omega_{a,b}n_a n_b\over 2} (c_2(L) - c_2(R)) + \sum_{a=1}^{r_{\mathfrak{g}}}  n_a \left(-\frac{1}{4}\text{Tr}F_{SU(2)_m}^2 + c_2(I)\right)\,,
\end{align}
with \(\Omega\) the Cartan matrix of the simply-laced Lie algebras \(\mathfrak{g}\) with rank \(r_{\mathfrak{g}}\). $c_2(L)$ and $c_2(R)$ the second Chern classes of two $SU(2)$ bundles that split the normal bundle of strings. Moreover, we have split the $SO(5)$ R-symmetry into $SU(2)_m \times SU(2)_R$ in \eqref{eq83_R4T2index}, and $F_{SU(2)_m}$ is the background field strength associated to the $SU(2)_m$ while $c_2(I)$ is the second Chern class of the $SU(2)_I$ bundle inherited from the $SU(2)_R$. 

$I_4$ is responsible for a non-trivial index. Indeed, if we perform an equivariant integration in the $\Omega$-background \cite{Bobev_2015} and get (we rescale the parameters to match our convention),
\begin{align}
\mathfrak{i}_{\mathfrak{n}}&=I_4\Bigr( c_2(L)\to -(\frac{\epsilon_-}{2\pi i})^2, \ c_2(R)\to -(\frac{\epsilon_+}{2\pi i})^2, \ \frac{1}{4}\text{Tr}F_{SU(2)_m}^2\to -(\frac{m}{2\pi i})^2,\ c_2(I) \to -(\frac{\epsilon_+}{2\pi i})^2  \Bigr)
\nonumber \\
&=-\frac{1}{4\pi^2}\left(-\sum_{a,b}{\Omega_{a,b}n_a n_b \over 2} (\epsilon_-^2-\epsilon_+^2)+\sum_a n_a(m^2-\epsilon_+^2)\right)\,,
\label{eq173_index}
\end{align}
then the elliptic genus has the following modular anomaly under an S-transformation \cite{DelZotto:2016pvm},
\begin{align}
Z_{\mathfrak{n}}(-{1\over \tau}|{\epsilon_{1,2}\over \tau},{m\over \tau}) = \exp\Bigr[2\pi i { \mathfrak{i}_\mathfrak{n} \over  \tau} \Bigr] \cdot Z_{\mathfrak{n}}(\tau|\epsilon_{1,2},m)\,.
\label{eq178_EG}
\end{align}
Comparing \eqref{eq178_EG} with the definition of Jacobi forms (\eqref{modtran} in the appendix \ref{App:MF}), we learn that $i_\mathfrak{n}$ is precisely its index.
\par 
Let us make a few comments about the existing literature. The 6d (2,0) A-type theories are given by the worldvolume theory on \(r_{\mathfrak{g}}+1\) M5-branes, and M2-branes ending on M5-branes become the self-dual strings in 6d. With this brane construction, one can engineer the 2d worldvolume theory on the \(\mathfrak{n}\) self-dual strings, which is given by the A-type quiver gauge theory with gauge group \(U(n_1)\times U(n_2)\times...\times U(n_{r_{\mathfrak{g}}})\) \cite{Haghighat:2013}.  The elliptic genus of an arbitrary base degree can be computed from the localization method. However, for the D/E-type SCFTs, we do not have closed-form expressions. In \cite{Gadde:2015tra, Haghighat:2017vch}, the strings in D/E-type theories were studied with the mass parameter \(m\) turned off. However, as far as we know, the fully refined elliptic genera have been remained to be unknown.

In the final part of this section, we talk about the connection to topological string theory. In order to do that, we first spell out some necessary ingredients involved. 

Given a CY threefold $X$, if we denote the (complexified) K\"{a}hler parameter of two cycles in $\alpha \in H_2(X, \mathbb{Z})$ as $t_\alpha$ with its exponential $Q_\alpha = \exp(i t_\alpha)$, the genus $g$ free energy $F_{g}$ has the following expansion,
\begin{equation}
    F_{g}(t) = \sum\limits_{\alpha \in H_2(X, \mathbb{Z})} r^{\alpha}_{g}\, \mathbf{Q}^{\alpha}\,.
\end{equation}
$r^{\alpha}_{g}$ are genus $g$ Gromov-Witten (GW) invariants, which are in general rational numbers. Furthermore, through lifting type IIA string theory to M-theory \cite{Gopakumar:1998ii, Gopakumar:1998jq, Gopakumar_1999}, we can do a partial resummation over $g_s$ to obtain another type of expansion, \footnote{We rescale the conventional definition of $g_s$ by $\sqrt{-1}$ to match the convention from gauge theory. Same as for $\epsilon_{\pm}$ in \eqref{TS:refinedGV}.}
\begin{equation}\label{TS:UnrefinedGV}
    F_{\text{top}}(t) = \sum\limits_{g \geq 0}\, F_{g}\, g_s^{2g-2} = \sum\limits_{g \geq 0} \sum\limits_{k \geq 1} \sum\limits_{\alpha \in H_2(X, \mathbb{Z})} (-1)^{g-1} I^{\alpha}_{g} \left(2 \sinh(k\frac{g_s}{2}) \right)^{2g - 2} \frac{\mathbf{Q}^{k \alpha}}{k}\,.
\end{equation}
Now $I^{\alpha}_{g}$ are always integers. They are related to M2-branes in the M-theory. From the 5d perspective, BPS M2-branes wrapped on two cycles $\alpha$ in $X$ give rise to BPS states in the spacetime. They are naturally labeled by two spins $j_-$ and $j_+$ indicating their representation under the little group $SO(4)$, and $I^{\alpha}_{g}$ is related to a Witten-like index over their Hilbert space ${\mathcal H}^{\alpha}_{\rm BPS}$,
\begin{equation}
    {\rm Tr}_{{\mathcal H}^{\alpha}_{\rm BPS}} (-1)^{2 j_+}{\mu}^{2 j^3_-} = \sum_{j_{\pm} \in \mathbb{N}/2} (-1)^{2 j_+} (2j_+ + 1) \chi_{j_-}(\mu)\,N^{\alpha}_{j_- j_+}  = \sum_{g=0}^{\infty} I^{\alpha}_{g} (\mu^{\frac{1}{2}} + \mu^{-\frac{1}{2}})^{2 g}\,.
\end{equation}
We introduce the symbol $\chi_{j}(x)$ to denote the following Laurent polynomial of an irreducible $SU(2)$ highest-weight representation with spin $j \in \mathbb{N}/2$,
\begin{equation}
    \chi_{j}(x) = x^{-2j} + x^{-2j+2}+ \cdots + x^{2j}\,,
\end{equation}
and $N^{\alpha}_{j_- j_+}$ is the number of multiplets for BPS states in ${\mathcal H}^{\alpha}_{\rm BPS}$ with spin $j_-$ and $j_+$. In general, as we move inside the moduli space of $X$, $N^{\alpha}_{j_- j_+}$ may vary, but $I^{\alpha}_{g}$ remains unchanged due to the property of the index. $I^{\alpha}_{g}$ are dubbed (unrefined) Gopakumar-Vafa (GV) invariants.

For local, i.e., non-compact CY threefolds, M-theory compactification engineers five dimensional gauge theory with $\mathcal{N} = 1$ supersymmetry. Putting the five dimensional spacetime under the $\Omega$-background, we have the Nekrasov partition function $Z(\epsilon_1,\epsilon_2,t)$ \cite{Nekrasov:2002qd,Nekrasov}, where $\epsilon_1$ and $\epsilon_2$ are two formal parameters associated to the Cartan subalgebra of $so(4) = su(2) \times su(2)$.

$Z(\epsilon_1,\epsilon_2,t)$ can be used to refine the topological string partition functions \cite{IKV,Aganagic:2011mi, Huang:2011qx}. It is worth emphasizing that this also holds for elliptically fibered CYs, which engineer five dimensional theories having six dimensional UV completions. The upshot is that the free energy enjoys a refined GV expansion,
\begin{equation}\label{TS:refinedGV}
    \begin{aligned}
    \mathcal{F}_{\text{ref}} &= \log \mathcal{Z}_{\text{ref}} = \log Z(\epsilon_1,\epsilon_2,t) \\
    & = \sum\limits_{2j_{\pm} \in \mathbb{N}} \sum\limits_{k \geq 1} \sum\limits_{\alpha \in H_2(X, \mathbb{Z})} N^{\alpha}_{j_-j_+} {(-1)^{2 (j_-+j_+)} \, \chi_{j_-}(u^k) \chi_{j_+}(v^k) \over v^k + v^{-k} - u^k - u^{-k}} \frac{\mathbf{Q}^{k \alpha}}{k}\,,
    \end{aligned}
\end{equation}
with $u = \exp(-\epsilon_-), v=\exp(-\epsilon_+)$ and $\epsilon_{\pm} = \frac{\epsilon_1 \pm \epsilon_2}{2}$.
Due to non-compactness, the $n^{\alpha}_{j_- j_+}$ defined earlier no longer depend on the moduli so are themselves well-defined quantities, known as refined GV invariants \cite{Hollowood:2003cv,IKV,Huang:2013}. Notice that they are always non-negative since they are counting numbers of BPS states.

The connection to topological string theory lies in the duality between F-theory and M-theory \cite{Klemm_1997}. It is known that F-theory compactification on $X \times S^1$ is dual to M-theory compactification on $X$. As discussed before, in the F-theory picture, BPS strings arise from D3-branes wrapping on curves $\Sigma_i$ inside $B$ and can further wrap on $T^2$ in the spacetime. While in the M-theory picture, they are dual to M2-branes wrapping on the same curves in $B$ and the elliptic fiber $T^2$ in $X$. The K\"{a}hler parameters of the rational curves are identified, while the radius of the extra $S^1$ in F-theory is inversely proportional to the volume of elliptic fiber in M-theory. Therefore, the fiber of $X$ not only has a complex structure $\tau$ but also acquires a (complexified) K\"{a}hler parameter $t$, which justifies the wrapping of M2-branes. 

After turning on the $\Omega$-background, the partition functions of self-dual BPS strings in F-theory get identified with the refined topological string partition function $\mathcal{Z}_{\text{ref}}$ of $X$ in  M-theory. To make the precise identification, in $Z_{\mathbb{R}^4\times T^2}$, we are only interested in the elliptic genera of BPS strings $Z_{T^2}$ \eqref{eq130_EGEXP}. While corresponding in $\mathcal{Z}_{\text{ref}}$ we are only summing over two-cycles that contains a non-trivial class in $H_{2}(B, \mathbb{Z})$. We abuse our notation and still call it $\mathcal{Z}_{\text{ref}}$, hoping that no confusion will occur,\footnote{In fact, the two $Z_{\mathfrak{n}}$ still differ by some overall factor in general \cite{Gu:2017ccq, DelZotto:2017mee}, which happens to be one for (2,0) theories.}
\begin{equation}\label{EG:expanZ}
    \mathcal{Z}_{\text{ref}} = 1 + \sum_{\mathfrak{n} \neq 0\, \in \mathbb{Z}_{\geq 0}^r} Z_{\mathfrak{n}}\, e^{-\mathfrak{n}\cdot \mathbf{v}}\,. 
\end{equation}
We summarize the duality in table \ref{Table:FMdual}.
\begin{table}[ht!]
\def\arraystretch{1.5}
    \centering
    \begin{tabu}{c|[1pt]c}\tabucline[1pt]{0-6}
        F$[X \times S^1]$\ & \ M$[X]$\\[-1pt] \tabucline[1pt]{0-6}
        $1/R_{S^1}$ & $\text{Vol}\ (T^2)$\\\hline
        $q = \exp(2\pi i\tau)$\ & \ $q = \exp(2\pi i t)$\\\hline
        D3-branes on $\Sigma_i$ and $T^2$ \ & M2-branes on $\Sigma_i$ and $T^2$\\\hline
        $\text{elliptic\ genus}: Z_{T^2}$ & $\text{partition function}: \mathcal{Z}_{\text{ref}}$\\ \tabucline[1pt]{0-6}
    \end{tabu}
    \caption{Duality between F-theory and M-theory.}
    \label{Table:FMdual}
\end{table}

Last but not least, the modular anomaly can also be motivated from topological string theory. This is not essential to the main result of this paper, and we include it here just for completeness. This part can be safely skipped for an uninterested reader. 

Above, what we already talked about is actually only the type-A topological string, and in fact, there exists a type-B topological string theory, based on coupling gravity to B-model on the worldsheet. In this setup, we also have a genus expansion of partition function, which now depends on the holomorphic structure rather than the K\"{a}hler structure of the CY manifold. At genus zero, the partition function is holomorphic. However, for the genus larger than zero, the anti-holomorphic part no longer decouples and gives rise to the so-called \textit{holomorphic anomaly} \cite{BCOVI,BCOV}. After a careful analysis of anti-holomorphic dependence, the authors in \cite{BCOVI,BCOV} found a set of recursive equations satisfied by the partition function, which are also known as the BCOV holomorphic anomaly equations. 

In \cite{Aganagic_2007}, it is pointed out that the holomorphic anomaly is tightly related to the modular anomaly. For cases of complex structure moduli space being one adimensional, this boils down to simply the choice of the quasimodular form $E_2$ versus almost holomorphic modular form $\hat E_2(\tau)$ (see appendix \ref{App:MF}) in a suitable parameterization. If we choose the quasimodular form $E_2$, the partition function will be holomorphic but no longer modular. 

For some classes of geometries that exhibit an elliptic fibration, \cite{HKK} proposed the following form of the anomaly equations,
\begin{equation}
    \left(\partial_{E_2} + \frac{1}{12} \frac{\mathfrak{n} \cdot (\mathfrak{n} + K_{B})}{2}\right)Z_{\mathfrak{n}} = 0\,,
\end{equation}
where the modular parameter of $E_2$ is identified with the complex parameter of the elliptic fiber, $\mathfrak{n}$ is now regarded as an element in $H_2(B, \mathbb{Z})$, $K_{B}$ means the canonical divisor of $B$ and the dot denotes intersection inside the base surface. In \cite{Gu:2017ccq}, this was generalized to the refined case with M-string mass turned on. In particular, for geometries related to ADE M-strings, $K_{B}$ is trivial, and they propose (we change it slightly to match our convention),
\begin{equation}\label{Mod_anomaly}
    \left(\partial_{E_2} + \frac{\pi^2}{3}\left[-\frac{1}{4\pi^2}\left(-\frac{\mathfrak{n}^2}{2}(\epsilon_+^2-\epsilon_-^2) - (\sum_{i} n_i)\epsilon_+^2 + (\sum_{i} n_i) m^2\right)\right]\right)Z_{\mathfrak{n}} = 0\,.
\end{equation}
By comparing \eqref{Mod_anomaly} and \eqref{App:DEJF}, we learn that \eqref{Mod_anomaly} motivates $Z_{\mathfrak{n}}$ to be a Jacobi form and the term inside the bracket is precisely its index. Since the curves inside the resolved ALE spaces intersect according to the negative of the corresponding Cartan matrix, we find a perfect agreement between topological string theory and gauge theory.
 
 \section{Elliptic genera of ADE M-strings}\label{sec:bootstrap}
 In this section, we obtain the elliptic genera of ADE M-strings using the modular bootstrap. In subsection \ref{sec:flip}, we will show that a few consistency conditions can determine the elliptic genera at low base degrees. In subsection \ref{section4}, we study the GV invariants and their vanishing bounds, which allow us to compute the elliptic genera at an arbitrary base degree.
 
\subsection{Modular bootstrap from the flip symmetry}\label{sec:flip}
\par 
In this subsection, we discuss some consistency conditions that the elliptic genus should follow, including the `flip symmetry,' which is a purely kinematical constraint for the self-dual strings in (2,0) theories. We will show that the elliptic genera of ADE-type M-strings can be bootstrapped with those consistency conditions up to some low base degrees. The extension to an arbitrary base degree will be discussed in subsection \ref{section4}.
\par 
As we have discussed in section \ref{section2}, the elliptic genus is a modular form with the modular parameter \(\tau\), and the elliptic parameters \(\epsilon_{1,2}\) and \(m\). Therefore, the elliptic genus of self-dual strings can be written as follows \cite{DelZotto:2016pvm},
\begin{align}
Z_{\mathfrak{n}}=\eta(\tau)^{n_0}\cdot 
{\mathcal{N}_{\mathfrak{n}}(\tau,m,\epsilon_{1,2}) 
\over \mathcal{D}_{\mathfrak{n}}(\tau,\epsilon_{1,2})}.
\label{eq219_ansatz}
\end{align}
Here \(\eta(\tau)\) is the Dedekind eta function,  and \(n_0\) is the weight of the elliptic genus which is 0 in our case. Also, \(\mathcal{N}\) is a numerator, and \(\mathcal{D}\) is a denominator of the elliptic genus which will be explained in the following paragraphs.
\par 
First, let us consider the denominator structure of the elliptic genus. In the thermodynamic limit \(\epsilon_{1,2}\to 0\), the free energy of the 6d \(\mathbb{R}^4\times T^2\) index should have the volume divergence as \(\log Z_{\mathbb{R}^4\times T^2}\propto {1\over \epsilon_1 \epsilon_2}\). Therefore, the elliptic genus also has a pole structure at \(\epsilon_{1,2}=0\). We take our denominator to capture all those volume divergences of \(\mathbb{R}^4\) of the elliptic genus, and it can be written in the following form \cite{DelZotto:2016pvm, Haghighat:2015ega,Gu:2017ccq},
\begin{align}\label{sec3:D}
\mathcal{D}_{\mathfrak{n}}=\prod_{a=1}^{r_{\mathfrak{g}} } \prod_{k=1}^{n_a} {\theta_1({k\epsilon_1\over 2\pi i})\over \eta(\tau)^3} {\theta_1({k\epsilon_2\over 2\pi i})\over \eta(\tau)^3}
\end{align}
where \(r_{\mathfrak{g}}\) is the rank of the Lie algebra \(\mathfrak{g}\) which is the type of the (2,0) SCFT. Note that the denominator itself is a modular form, and it transforms as follows under S-duality,
\begin{align}
\mathcal{D}_{\mathfrak{n}}(-{1\over \tau},{\epsilon_{1,2}\over \tau})
=\tau^{-2\sum_a n_a} \exp\Bigr[{1\over 4\pi i\tau}\sum_{a=1}^{r_{\mathfrak{g}}}
\sum_{k=1}^{n_a} k^2 (\epsilon_1^2+\epsilon_2^2) \Bigr]
\cdot \mathcal{D}_{\mathfrak{n}}(\tau,\epsilon_{1,2})
\end{align}
which means that the denominator has the weight \(w_{\mathcal{D}}=-2\sum_a n_a\) and the index \(i_{\mathcal{D}}={1\over 2 (2\pi i)^2}\sum_{a=1}^{r_{\mathfrak{g}}} \sum_{k=1}^{n_a} k^2(\epsilon_1^2+\epsilon_2^2)\).
\par 
Now, let us move on to the structure of the numerator \(\mathcal{N}_\mathfrak{n}\). Note that it depends on the elliptic parameters \(m\) and \(\epsilon_{1,2}\), which are also the chemical potentials of the \(SU(2)\) global symmetries. Therefore, one can capture the dependence of \(m\) and \(\epsilon_\pm\) in the elliptic genus in terms of the \(SU(2)\) Weyl-invariant Jacobi forms: \(\phi_{0,1}(\tau,z)\) and \(\phi_{-2,1}(\tau,z)\). They are the modular forms which are invariant under the \(SU(2)\) Weyl transformation of \(z\), and \(\phi_{k,l}(\tau,z)\) has an index \(l z^2\) and weight \(k\). See the appendix \ref{appendixJMF} for detailed information. Then, one can write down the numerator in the following form,
\begin{align}
\mathcal{N}_{\mathfrak{n}}=\sum_{\mathfrak{p} }C_{\mathfrak{p}}E_4(\tau)^{p_4} E_6(\tau)^{p_6} \prod_{i=1}^3 [\varphi_{-2,1}(\tau,{z_i\over 2\pi i })]^{p_{-2,i}} [\varphi_{0,1}(\tau,{z_i\over 2\pi i})]^{p_{0,i}}.
\label{eq115_numerator}
\end{align}
Here, \(E_{4,6}(\tau)\) are the Eisenstein series, and chemical potentials are written as \((z_1,z_2,z_3)=(\epsilon_+,\epsilon_-,m)\). Note that we express the numerator in terms of the Eisenstein series and the \(SU(2)\)-invariant Jacobi forms. The coefficient is denoted as \(C_{\mathfrak{p}}\) which depends on the vector \(\mathfrak{p}=(p_4,p_6,p_{-2,1},p_{0,1},p_{-2,2},p_{0,2},p_{-2,3},p_{0,3}) \in (\mathbb{Z}_{\geq 0})^8\).  \eqref{sec3:D} and \eqref{eq115_numerator} is precisely our modular ansatz for $Z_\mathfrak{n}$.
\par 
The elliptic genus (\ref{eq219_ansatz}) is determined once we can fully fix the coefficient \(C_\mathfrak{p}\) of the numerator (\ref{eq115_numerator}). Fortunately, even without knowing the 2d CFT on the self-dual string, there are some consistency conditions for the elliptic genus that we can use to fix the numerator. First, one can use the modular property (\ref{eq173_index}) of the elliptic genus. Then, the index and the weight of the numerator is given as follows,
\begin{align}
w_{\mathcal{N}}&=-2\sum_{a=1}^{r_{\mathfrak{g}} } n_a
\nonumber \\
i_{\mathcal{N}}&={1\over 2(2\pi i)^2}\Bigr(\epsilon_1 \epsilon_2 \sum_{a,b}^{r_G} \Omega_{a,b}n_a n_b +2(m^2-\epsilon_+^2)\sum_{a=1}^{r_{\mathfrak{g}} } n_a
+\sum_{a=1}^{r_{\mathfrak{g}} } \sum_{k=1}^{n_a}k^2(\epsilon_1^2+\epsilon_2^2) \Bigr).
\label{eq107_indexweight}
\end{align}
From (\ref{eq107_indexweight}), it is straightforwards to see that the power vector \(\mathfrak{p}\) should satisfy the following two relations,
\begin{align}
w_{\mathcal{N}}&=-2\sum_{i=1}^3 p_{-2,i}+4p_4+6p_6
,\quad 
i_{\mathcal{N}}=\sum_{i=1}^3 (p_{-2,i}+p_{0,i})({z_i\over 2\pi i})^2.
\label{eq119_wslwid}
\end{align}
Since all \(p\)'s are non-negative integers, there are only a finite number of solutions for (\ref{eq119_wslwid}). All the coefficient \(C_{\mathfrak{p}}\)'s are zero if \(\mathfrak{p}\) does not satisfy (\ref{eq119_wslwid}). The remaining problem is to determine the finite number of the allowed coefficients.
\par 
One way to determine those non-zero coefficients is to use the instanton expansion of the 5d maximally supersymmetric Yang-Mills theory (MSYM) with gauge group \(G\), which is the circle compactified 6d (2,0) \(\mathfrak{g}\)-type theory. Until now, we considered the elliptic genus expansion (\ref{eq130_EGEXP}) of the 6d \(\mathbb{R}^4\times T^2\) index,
\begin{align}
Z_{\mathbb{R}^4\times T^2}(\tau,m,\epsilon_\pm,\mathbf{v})=Z_0\Bigr(1+ \sum_{\mathfrak{n}}Z_{\mathfrak{n}}(\tau,m,\epsilon_\pm)e^{-\mathfrak{n}\cdot \mathbf{v}}\Bigr)
\end{align}
whose expansion parameter is the string fugacity \(e^{-\mathbf{v}}\) and the momentum mode \(Z_0\) is given as follows,
\begin{equation}
    Z_0=\text{PE}\Bigr[r_{\mathfrak{g}} {\sinh{m\pm \epsilon_-\over 2}\over \sinh{\epsilon_{1,2}\over 2}} {q\over 1-q}\Bigr].
    \label{eq280_Z0}
\end{equation}
Here, PE stands for the plethystic exponential, defined as
\begin{equation}
    \text{PE}[f(x_1,x_2,\cdots,x_n)] := \exp\left(\sum_{k = 1}^{\infty}\frac{1}{k} f(x_1^k, x_2^k, \cdots, x_n^k)\right)
\end{equation}
with \(x\)'s are fugacity-like variables. Instead, one can consider a different form of the expansion as follows,
\begin{align}
Z_{\mathbb{R}^4\times T^2}(\tau,m,\epsilon_\pm,\mathbf{v})=Z_\text{pert}
\Bigr(1+\sum_{k=1}^\infty Z_k(m,\epsilon_\pm,\mathbf{v}) q^k\Bigr)
\label{eq286_INST}
\end{align}
where the expansion parameter is the instanton fugacity \(q=e^{2\pi i \tau}\). In the above formulation, one views the 6d \(\mathfrak{g}\)-type (2,0) SCFTs on \(\mathbb{R}^4\times T^2\) as the 5d MSYM on \(\mathbb{R}^4\times S^1\) with gauge group \(G\). Here, the KK spectrum of the compactified circle becomes the instantons in 5d \cite{Kim:2011mv}. \(Z_\text{pert}\) is the perturbative partition function of the 5d MSYM given as follows \cite{Bullimore:2014awa},
\begin{align}
Z_\text{pert}=\text{PE}\Bigr[{\sinh{m\pm \epsilon_+\over 2}\over \sinh{\epsilon_{1,2}\over 2}} \sum_{\alpha\in\Psi^+_{\mathfrak{g}} } e^{-\alpha(\mathbf{v})} \Bigr]
\end{align}
where \(\Psi^+_{\mathfrak{g}}\) is a set of positive roots of the Lie algebra \(\mathfrak{g}\). Also, \(Z_k\) is the \(k\)-instanton partition function. It can be computed from ADHM quantum mechanics when \(G\) is A- or D-type \cite{Hwang:2014uwa, Hwang:2016gfw}. However, when \(G\) is E-type, there are no known results for the 5d MSYM instanton partition function. Since we are interested in the elliptic genus of the self-dual strings in generic 6d (2,0) SCFTs including the exceptional Lie algebras, we shall not use the instanton partition function to determine the elliptic genus. Instead, we shall use the perturbative data only, by imposing the following condition,
\begin{align}
[Z_0 Z_{\mathfrak{n}}(\tau,m,\epsilon_\pm) ]_{q^0} = [Z_\text{pert}(m,\epsilon_\pm ,\mathbf{v})]_{e^{-\mathfrak{n}\cdot \mathbf{v}}}.
\label{eq270_PertMatching}
\end{align}
Here \([f]_q\) denotes the coefficient of \(q^0\) of \(f\), and \([f]_{e^{-\mathfrak{n}\cdot \mathbf{v}}}\) denotes the coefficient of \(e^{-\mathfrak{n}\cdot \mathbf{v}}\) of \(f\). 
\par 
Lastly, there is another condition, which we will call a `flip symmetry,' that can be used to fix the elliptic genus. The flip symmetry is defined as the invariance of the elliptic genus under the following flip transformation,
\begin{align}
e^{-\epsilon_+}\to -e^{-\epsilon_+},\quad 
e^{-\epsilon_-}\to -e^{-\epsilon_-},\quad 
e^{-m}\to -e^{-m}.
\label{eq279_transofrm}
\end{align}
Under the flip transformation, the elliptic genus changes as follows,
\begin{align}
Z_{\mathfrak{n}}&=\text{Tr}[(-1)^F q^P e^{-\epsilon_1 J_1-\epsilon_2 J_2}
e^{-(\epsilon_++m)Q_1-(\epsilon_+-m)Q_2}]
\nonumber \\
&\to 
\text{Tr}[e^{-2\pi i (J_1+Q_1)}\cdot (-1)^F q^P e^{-\epsilon_1 J_1-\epsilon_2 J_2}
e^{-(\epsilon_++m)Q_1-(\epsilon_+-m)Q_2}].
\end{align}
Now, recall that \(J_{1,2}\) are the spins of the \(SO(4)\) Lorentz symmetry of the tangent \(\mathbb{R}^4\), and \(Q_{1,2}\) are the spins of the unbroken \(SO(4)\) R-symmetry of the normal \(\mathbb{R}^4\). Therefore, \(J_1\) and \(Q_1\) are integers for bosons and half-integers for fermions, i.e, \(e^{-2\pi i J_1}=e^{-2\pi i Q_1}=(-1)^F\).\footnote{This relation also plays an important role in the study of the modified indices \cite{Choi:2018hmj, Kim:2019yrz, Nahmgoong:2019hko}.} Then, it is trivial to check that \(e^{-2\pi i (J_1+Q_1)}=1\) and the elliptic genus is invariant under the transformation (\ref{eq279_transofrm})
\par 
\begin{table}[t!]
\centering
\begin{tabular}{|c|c|c|c|}\hline
Base degree & Allowed terms & Perturbative data & Flip symmetry \\ \hline \hline
\(\begin{pmatrix}
2
\end{pmatrix}\) & 44 & 38 & 6 \\ \hline 
\(\begin{pmatrix}
2&1
\end{pmatrix}\) & 103 & 89 & 14 \\ \hline 
\(\begin{pmatrix}
1&2&1
\end{pmatrix}\) & 165 & 140 & 25 \\ \hline 
\(\begin{pmatrix}
&1& \\
1&2&1
\end{pmatrix}\) & 221 & 181 & 40 \\ \hline 
\(\begin{pmatrix}
&&1& \\
1&1&2&1
\end{pmatrix}\) & 292 & 231 & 61 \\ \hline 
\(\begin{pmatrix}
&&1&& \\
1&1&2&1&1
\end{pmatrix}\) & 374 & 286 & 88 \\ \hline 
\(\begin{pmatrix}
&&1&&& \\
1&1&2&1&1&1
\end{pmatrix}\) & 468 & 346 & 122 \\ \hline 
\(\begin{pmatrix}
&&1&&&&\\
1&1&2&1&1&1&1
\end{pmatrix}\) & 575 & 411  & 164 \\ \hline 
\end{tabular}
\caption{The `allowed terms' column denotes the number of solutions of the modular constraints (\ref{eq107_indexweight}). The `perturbative data' column denotes the number of the terms whose coefficients can be fixed by the perturbative data using (\ref{eq270_PertMatching}). The `flip symmetry' column denotes the number of the terms whose coefficients can be further fixed by the flip symmetry using (\ref{eq279_transofrm}). }
\label{tab370_FLIP}
\end{table}
The flip symmetry is a purely kinematic constraint in the sense that it does not depend on the microscopic details of the 2d CFT on the self-dual strings. However, it provides strong constraints when we fix the elliptic genus with the modular bootstrap. Surprisingly, the elliptic genera up to some low base degree can be completely determined from the conditions that we have mentioned so far. More precisely, the elliptic genus of the self-dual strings in 6d (2,0) SCFT can be fully determined up to some base degrees from the three conditions: {\textit {the modular property}}  (\ref{eq119_wslwid}), {\textit{the 5d perturbative data}} (\ref{eq270_PertMatching}), and {\textit{the flip symmetry}} (\ref{eq279_transofrm}). In table \ref{tab370_FLIP}, we tabulated the number of the coefficients of some elliptic genera whose coefficients can be totally fixed by the aforementioned three conditions. The full expression of the elliptic genera can be found in appendix \ref{append_EG}.
\par 
Although the flip symmetry can give non-trivial results for the elliptic genera, it does not give us the answer up to an arbitrarily high base degree. In the next subsection, based on the data obtained from the flip symmetry, we will establish a more powerful method that enables us to compute the elliptic genus at an arbitrary base degree.

\subsection{Modular bootstrap from the vanishing bound}
\label{section4}
In this subsection, we first argue the existence of vanishing bound for the GV invariants. Then we present our conjecture on such a bound based on the data at low base degrees. Together with the modular ansatz introduced in the previous subsection, it can recursively bootstrap all the elliptic genera of ADE-type (2,0) theories in principle. Finally, we address the issue of uniqueness for our solutions from modular bootstrap and make a comparison with the approach in subsection \ref{sec:flip}.

As mentioned in the introduction, a very interesting feature for the refined GV invariants $N^{\alpha}_{j_-j_+}$ is its vanishing property. But to understand that, let us first start from unrefined GV invariants $I_{g}^{\alpha}$. We would like to first show that for a fixed curve class $\alpha \in H_2(X ,\mathbb{Z})$, $I_{g}^{\alpha}$ all vanish when $g$ is sufficiently large. If the CY manifold is a total space of canonical line bundle over a complex surface, this can be argued for using the adjunction formula \cite{Gu:2017ccq}. However, the geometries we consider in this paper do not belong to that class, so we need a more general argument. Below, we sketch how the vanishing condition can be shown.\footnote{Another possible way of reasoning is to use the geometric model developed in \cite{KKV}.}

In the mathematical literature, there is a class of closely related enumerative invariants called stable pair or Pandharipande-Thomas (PT) invariants $P_{n,\alpha}$ \cite{PT1}. The starting point is to consider a torsion sheaf $F$ having dimension one support inside $X$, together with a non-trivial holomorphic section $s$ over it. We impose certain stability conditions, such that the zeros of sections are only a bunch of points. Physically speaking, the one-dimensional support corresponds to a D2-brane, and the points correspond to D0-branes. Therefore, a stable pair $(F, s)$ can be roughly thought of as a D0-D2 brane system. 

The set of all possible stable pairs with a fixed holomorphic Euler characteristic $n$ and a fixed class of support $\alpha$ naturally forms a moduli space $P_n(X,\alpha)$. The existence of a virtual fundamental cycle was shown \cite{PT1}, which happens to be zero-dimensional for any CY threefold. Then the stable pair or PT invariants $P_{n,\alpha}$ just count the number of certain points inside $P_n(X,\alpha)$. The vanishing of $P_{n,\alpha}$ for a fixed $\alpha$ and small enough $n$ can be easily explained because a curve with a given homology class can not have arbitrarily small holomorphic Euler characteristic, rendering the moduli space empty.

On the other hand, $P_{n,\alpha}$ can be expressed as combinations of $I_{g}^{\alpha}$ \cite{PT1}. For instance, for an irreducible class $\beta \in H_2(X,\mathbb{Z})$, they are related in terms of formal power series in $q$,
\begin{equation}
    \sum_{n} P_{n,\beta}\, q^n = \sum_{g \geq 0} I_{g}^{\beta} q^{1-g} (q + 1)^{2g-2}\,.
\end{equation}
It turns out that to satisfy the vanishing property of $P_{n,\beta}$, $I_{g}^{\beta}$ must also vanish when $g$ is sufficiently large. Then we argue inductively and extend the vanishing result for $I_{g}^{\beta}$ to all curve classes in $H_2(X,\mathbb{Z})$. In contrast, the GW invariants do not have such a nice property due to the presence of "bubbling phenomenon" \cite{Katz_1996}.

As for the refined GV invariants, the geometric model is more complicated so it is harder to argue from the geometry. Instead, we adopt a different strategy, through exploring the relation between the refined GV invariants and the unrefined GV invariants.

First, we need to make an important assumption: The refined GV invariants satisfy a so-called \textit{checkerboard} pattern: fix a given curve class $\alpha$ and tabulate all non-zero invariants with two axes $2j_+$ and $2j_-$. Then any two occupied blocks are either disconnected or connected through a diagonal, which makes the table resemble a checkerboard. This pattern was first noticed in \cite{Choi:2012jz} and holds true for all non-compact CY threefolds that the authors know of. This also applies to the geometries considered in this paper, and we refer readers to appendix \ref{appen:BPS} for examples.

Assuming the above, we next compare \eqref{TS:UnrefinedGV} and \eqref{TS:refinedGV}. For the reader's convenience, we repeat them below,
\begin{equation}\label{section4:compare}
    \begin{aligned}
    &F_{\text{top}}(t) = \sum\limits_{g \geq 0} \sum\limits_{k \geq 1} \sum\limits_{\alpha \in H_2(X, \mathbb{Z})} (-1)^{g-1} I^{\alpha}_{g} \left(2 \sinh(k\frac{g_s}{2}) \right)^{2g - 2} \frac{\mathbf{Q}^{k \alpha}}{k}\,,\\
    \mathcal{F}_{\text{ref}} & = \sum\limits_{2j_{\pm} \in \mathbb{N}} \sum\limits_{k \geq 1} \sum\limits_{\alpha \in H_2(X, \mathbb{Z})} N^{\alpha}_{j_-j_+} {(-1)^{2 (j_-+j_+)} \, \chi_{j_-}(u^k) \chi_{j_+}(v^k) \over v^k + v^{-k} - u^k - u^{-k}} \frac{\mathbf{Q}^{k \alpha}}{k}\,.
    \end{aligned}
\end{equation}
The unrefined limit is $\epsilon_1 = - \epsilon_2 = g_s$, from which we can deduce the following,
\begin{equation}
    \sum\limits_{2j_{\pm} \in \mathbb{N}} N^{\alpha}_{j_-j_+} (-1)^{2 (j_-+j_+)} (2j_+ + 1)\, \chi_{j_-}(\text{e}^{- k g_s}) = \sum\limits_{g} I^{\alpha}_{g} (-1)^{g}\left(2 \sinh(k \frac{g_s}{2}) \right)^{2g}\,,
\end{equation}
for given $\alpha \in H_2(X, \mathbb{Z})$ and $k \in \mathbb{Z}_+$.

Thanks to the checkerboard pattern, $\left(2j_-+2j_+\right)$ is always even or odd for a fixed curve class $\alpha$. Moreover, $\chi_{j_-}(\text{e}^{- k g_s})$ can be decomposed into a polynomial with variable $t = -\left(2 \sinh(k \frac{g_s}{2}) \right)^{2}$,
\begin{equation}
    \chi_{j_-}(\text{e}^{- k g_s}) = a_{2j_-} t^{2j_-} + a_{2j_- - 1} t^{2j_- - 1} + \cdots + a_{0}\,.
\end{equation}
It is easy to show that $a_{k} > 0$ for $k$ even and $a_{k} < 0$ for $k$ odd always, regardless of $j_-$. Recall that $N^{\alpha}_{j_-j_+}$ is always a non-negative integer, this means for a given genus $g$, $I^{\alpha}_{g}$ can be written as a sum of non-zero $N^{\alpha}_{j_-j_+}$ with $2j_- \geq g$, and each term has the same sign.

As a consequence, there must exist vanishing bound for $j_{+}$ since $I^{\alpha}_{g}$ is a finite number, and also for $j_{-}$, since there is a vanishing bound for $g$ in the unrefined case. As a by-product, this explains the pattern of alternating signs of unrefined GV invariants when we increase the genus.\footnote{See appendix \ref{appen:BPS} for some examples.}

However, the bound for $j_{\pm}$ beyond which refined GV invariants all vanish depends on the details of geometry. Geometries involved here are elliptic fibrations over ALE spaces with mass deformations thereof. In practice, there are two issues. The first one is that it is not clear to us how to construct a compact embedding of $X$ into an ambient toric variety, perhaps except for the $A_r$ case. This prevents us from using many powerful techniques such as mirror symmetry. The other is that how to mass deform the geometry away from the trivial fibration is not known, except for the $A_1$ case \cite{gu2019elliptic}. For some relevant discussions, see \cite{gu2019elliptic, Gu:2020fem}. Last but not least, even if one finds ways to overcome the two issues, it is perhaps equally hard to show a useful vanishing condition based on the geometry.

In short, we are far away from being able to derive an optimal vanishing bound from first principles. So instead, we guess an empirical bound for the GV invariants based on the data at low base degrees obtained from the "flip symmetry" method in subsection \ref{sec:flip}. The bound is not complicated, so it doesn't saturate for many cases, but we believe that it is efficient enough for practical use.

To state our conjecture, we first fix the notation. Specialize the general form \eqref{TS:refinedGV} or \eqref{section4:compare} to our case,
\begin{equation}
    \mathcal{F}_{\text{ref}} = \sum\limits_{2j_{\pm} \in \mathbb{N}} \sum\limits_{k \geq 1} \sum\limits_{\mathfrak{n},e,\mu} \frac{N_{j_-j_+}^{\,\mathfrak{n},e,\mu}}{k} {(-1)^{2 (j_-+j_+)} \, \chi_{j_-}(u^k) \chi_{j_+}(v^k) \over v^k + v^{-k} - u^k - u^{-k}} \mathbf{Q}^{k \mathfrak{n}}\, q^{ke}\, Q_m^{\, k\mu}\,,
\end{equation}
where $\mathbf{Q}=e^{-\mathbf{v}}$, $q=e^{2\pi i \tau}$ and $Q_m=e^{-m}$ are formal exponential of K\"{a}hler parameters associated to exceptional curves in $B$, elliptic fiber and two-cycle of M-string mass respectively. In this representation $\mathfrak{n}$ and $e$ are non-negative, but $\mu$ also takes negative values. It is possible to redefine $q$ to get rid of negative $\mu$, such that the expansion is convergent at large radius limit. However, we choose to use the present form, in order to have the symmetry $\mu \rightarrow -\mu$ in $\mathcal{F}_{\text{ref}}$. Then due to $N_{j_-j_+}^{\,\mathfrak{n},e,\mu} = N_{j_-j_+}^{\,\mathfrak{n},e,-\mu}$, we only need to focus on $\mu$ being non-negative. Our conjecture for the vanishing bound is the following.
\begin{conj}
For elliptically fibered CY threefolds that engineer six-dimensional $\mathcal{N} = (2,0)$ ADE-type superconformal theories, their refined GV invariants satisfy uniformly the following vanishing bound:
\begin{equation}\label{Vanish:bound}
N_{j_-j_+}^{\,\mathfrak{n},e,\mu \geq 0}=0\ {\rm for}\ \left\{\begin{array}{rl} 2 j_- &> e \cdot \max \{n_i\} - \frac{\mu-e-1}{2} \cdot H( \mu - e -2)\\ & 
		{\rm \ \ \ \ \ \ \ \ \ \ \ \ \ \ \ \ or}  \\   
		2 j_+&> (e + 2) \cdot \max \{n_i\} - \frac{\mu - e - 1}{2} \cdot H( \mu - e -2)
	\end{array} \right.
	\end{equation}
\end{conj}
On the right-hand side, $H(x)$ is the Heaviside step function,
\begin{equation}
    \begin{aligned}
        H(x) = \begin{cases}0, & \text{if}\ x < 0 \\
      1, & \text{otherwise}
    \end{cases}
    \end{aligned}
\end{equation}

We explain briefly how to use the above constraint to determine the partition function or elliptic genus recursively. This starts from the relation between the partition function and the free energy,
\begin{equation}\label{vanishing:FtoZ}
    \mathcal{Z}_{\text{ref}} = 1 + \sum_{\mathfrak{n} \neq 0\, \in \mathbb{Z}_{\geq 0}^r} Z_{\mathfrak{n}}\, \mathbf{Q}^{\mathfrak{n}} =  \exp \left(\mathcal{F}_{\text{ref}}\right) = \exp\left(\sum_{\mathfrak{n} \neq 0\, \in \mathbb{Z}_{\geq 0}^r} F_{\mathfrak{n}} \mathbf{Q}^{\mathfrak{n}}\right)\,,
\end{equation}
where in the first equality we use \eqref{EG:expanZ}, and the second equality holds after we Taylor expand the exponential in terms of $\mathbf{Q}$. Therefore, for a given $\mathfrak{n}_{\alpha} = \{n_1,\cdots,n_r\}$ we have,
\begin{equation}
    Z_{\mathfrak{n}_{\alpha}} = F_{\mathfrak{n}_\alpha} + \sum_{j=2}^\infty
  \frac{1}{j!} \sum_{\mathfrak{n}_1, \ldots, \mathfrak{n}_j > \boldsymbol{0} \atop \sum \mathfrak{n}_i = \mathfrak{n}_{\alpha}}
  \prod_{i=1}^j F_{\mathfrak{n}_i}\,.
\end{equation}
Clearly, the second term in the right-hand side only involves $F_\mathfrak{n}$ of base degrees strictly smaller than $\mathfrak{n}_{\alpha}$. To have some feeling of this equation, we work out the expansion of the first few degrees for the E6-type,
\begin{equation}
\begin{aligned}
    Z_{\{1,0,0,0,0,0\}} &= F_{\{1,0,0,0,0,0\}}\,,\\
    Z_{\{1,1,0,0,0,0\}} &= F_{\{1,1,0,0,0,0\}} + \left(F_{\{1,0,0,0,0,0\}}\cdot F_{\{0,1,0,0,0,0\}}\right)\,,\\
    Z_{\{2,1,0,0,0,0\}} &= F_{\{2,1,0,0,0,0\}} + \big(F_{\{1,1,0,0,0,0\}}\cdot F_{\{1,0,0,0,0,0\}} + F_{\{2,0,0,0,0,0\}}\cdot F_{\{0,1,0,0,0,0\}}\big)\\
    &+ \big(\frac{1}{2}F_{\{1,0,0,0,0,0\}}^2\cdot F_{\{0,1,0,0,0,0\}}\big)\,.\\
\end{aligned}
\end{equation}
For our purpose, it is better to invert the above relations,
\begin{equation}
\begin{aligned}
    F_{\{1,0,0,0,0,0\}} &= Z_{\{1,0,0,0,0,0\}}\,,\\
    F_{\{1,1,0,0,0,0\}} &= Z_{\{1,1,0,0,0,0\}} - \left(Z_{\{1,0,0,0,0,0\}}\cdot Z_{\{0,1,0,0,0,0\}}\right)\,,\\
    F_{\{2,1,0,0,0,0\}} &= Z_{\{2,1,0,0,0,0\}} - \big(Z_{\{1,1,0,0,0,0\}}\cdot Z_{\{1,0,0,0,0,0\}} + Z_{\{2,0,0,0,0,0\}}\cdot Z_{\{0,1,0,0,0,0\}}\big)\\
    &+ \big(Z_{\{1,0,0,0,0,0\}}^2\cdot Z_{\{0,1,0,0,0,0\}}\big)\,.\\
\end{aligned}
\end{equation}
In general we can take the logarithm of both sides of \eqref{vanishing:FtoZ},
\begin{equation}\label{Vanishing:Boot}
  F_{\mathfrak{n}_\alpha} = Z_{\mathfrak{n}_\alpha} + \sum_{j=2}^\infty
  \frac{1}{j!}
  \sum_{\mathfrak{n}_1, \ldots, \mathfrak{n}_j > \boldsymbol{0} \atop\sum \mathfrak{n}_i = \mathfrak{n}_{\alpha}}
  a^{(\mathfrak{n}_\alpha)}_{\mathfrak{n}_1,\ldots,\mathfrak{n}_j} \prod_{i=1}^j
  Z_{\mathfrak{n}_i}\,,
\end{equation}
with $a^{(\mathfrak{n}_\alpha)}_{\mathfrak{n}_1,\ldots,\mathfrak{n}_j}$ some integer coefficients that can be determined inductively. Clearly, the second term in the right-hand side only involves $Z_\mathfrak{n}$ of base degrees strictly smaller than $\mathfrak{n}_{\alpha}$.

Our strategy goes as follows. We choose $\mathfrak{n}_{\alpha}$ small enough such that all the partition functions with degree lower than $\mathfrak{n}_{\alpha}$ are known, e.g., from the "flip symmetry" method in subsection \ref{sec:flip}. Then we impose both the vanishing condition in the left-hand side and the modular ansatz in the right-hand side of \eqref{Vanishing:Boot}. This turns to give enough constraints to completely fix $F_{\mathfrak{n}_\alpha}$ hence $Z_{\mathfrak{n}_{\alpha}}$. Increasing the degree one step at each time, we are able to determine the elliptic genus for any value of $\mathfrak{n}_{\alpha}$ in principle.

There is one important issue that needs careful discussion. Suppose we find one consistent solution to the above equation, how can we guarantee that it corresponds to the solution we want? In other words, can we show that the solution is unique? Note that if the solution is unique, then the vanishing bound must be able to fix all the coefficients in the ansatz. Happily, this point was already discussed in depth in \cite{Gu:2017ccq}. By invoking the GV expansion and the fact that all weak Jacobi forms have a non-negative index, the authors show the following criterion \cite{Gu:2017ccq}: 

\begin{thm}\label{Vanishing:thm}
If the index of either $\epsilon_+$ or $\epsilon_-$ for $Z_{\mathfrak{n}_{\alpha}}$ is smaller than $-1$, then the solution to \eqref{Vanishing:Boot} must be unique.
\end{thm}

Let us apply it to our situation. It so happens that the index of $\epsilon_-$ takes the form,
\begin{equation}
    i_{\epsilon_-} = -\frac{1}{2} \mathfrak{n}\cdot\Omega\cdot\mathfrak{n}^{\text{T}}\,,
\end{equation}
where $\Omega$ is again the Cartan matrix. If we embed the simple roots into the Euclidean space, then $\Omega$ will map to the standard Euclidean inner-product and $-2i_{\epsilon_-}$ the norm-square of a given root. Since they are all even integral lattices, the smallest non-zero possible value for length-square is two, and the next one is four. Thus based on the criterion \ref{Vanishing:thm}, only non-zero vectors with the shortest length may cause issues. For simply-laced Lie algebras, they are nothing but the positive roots. Although this could be treated generally, let us just do a case-by-case analysis here. For $\mathfrak{a}_1$, $\mathfrak{a}_2$, $\mathfrak{a}_3$, $\mathfrak{d}_4$, $\mathfrak{d}_5$, $\mathfrak{e}_6$, $\mathfrak{e}_7$ and $\mathfrak{e}_8$, there are 1, 3, 6, 12, 20, 36, 63 and 120 in total respectively, and it is straightforward to enumerate them on a computer. Then we need to determine the index of $i_{\epsilon_+}$for those vectors. It turns out that in each case, for most of them, the index $i_{\epsilon_+}$ is, fortunately, smaller than $-1$, and all exceptional vectors are from low enough base degrees such that their free energies can be unambiguously determined by the "flip symmetry" method in subsection \ref{sec:flip}.\footnote{For example, they are not larger than the case where the degrees wrapping exceptional cycles are all equal to one.}

To summarize, combining information from both gauge theories and topological strings, we are able to show that our bootstrapping procedure gives the correct answer for the elliptic genera, provided that the conjecture \ref{Vanish:bound} is valid.

The maximal degrees that we have computed for each theory are listed in table \ref{tabEGA1} - \ref{tabEGE8} for \(A_{1,2,3}\), \(D_{4,5}\), and \(E_{6,7,8}\)-type theories. As a consistency check, one can compare our bootstrapped elliptic genera with the known results. For A-type theories, its localization method was studied in \cite{Haghighat:2013}, where the formula is given. We checked that the bootstrap gives exactly the same results for all the A-type elliptic genera we computed in table \ref{tabEGA1}. For D/E-type theories, its elliptic genera is currently unknown unless \(m\) is turned off \cite{Gadde:2015tra, Haghighat:2017vch}. Instead, one can use the instanton partition function of 5d D-type MSYM to test our elliptic genus. We checked that the bootstrap gives exactly the same results with the instanton partition function up to \(\mathcal{O}(q^1)\) for all the D-type elliptic genera we computed in table \ref{tabEGD4} and \ref{tabEGD5}. Especially for the \(D_4\) case, the elliptic genera of some low base degrees are already computed in \cite{Gu:2017ccq, Kim:2018gak}. We also confirmed that our results are the same. 
\par
Lastly, for the E-type cases, there have been no known fully refined results before this paper. Therefore, we would like to emphasize that the results given in table \ref{tabEGE6}, \ref{tabEGE7}, and \ref{tabEGE8} are the first non-perturbative computations for the fully refined elliptic genera of the 6d (2,0) E-type SCFTs.
\par 
As a final remark, let's make a comparison between the "flip symmetry" method in subsection \ref{sec:flip} and the "vanishing bound" method in this subsection. On the one hand, they both rely on the modular ansatz for the partition function to start with. On the other hand, although the flip symmetry is only a kinematical constraint, it can be imposed directly on the partition function itself. Together with the perturbative part, it usually fixes most, if not all, of the coefficients in the modular ansatz at low base degrees. The rest can possibly be determined if we further demand, e.g., the GV form for the free energy, instanton partition functions expansion if available. On the other hand, the vanishing bound for GV invariants is more powerful, albeit conjectural. To make use of it, we need to subtract all the previous partition functions of smaller base degrees, which often becomes quite time-consuming in practice. This explains why we could not go very far beyond the cases already covered by the "flip symmetry" method. Also, its efficiency depends crucially on the precision of the vanishing bound.

\section{Applications}\label{section5}
This section discusses the two applications of the elliptic genera that we obtained from the modular bootstrap. We will mainly focus on the generalization of the results previously known only for A or D-type to all ADE theories. First, we shall compute the 6d (2,0) Cardy formulas \cite{Lee:2020rns} on \(\mathbb{R}^4\times T^2\) to all ADE-type. Second, we shall compute the (2,0) superconformal index and show that it becomes the \(\mathcal{W}\)-algebra character in the unrefined limit for all ADE-type.

\subsection{Cardy limits on \(\mathbb{R}^4\times T^2\)}
In this subsection, we compute the Cardy formulas of general 6d (2,0) SCFTs on \(\mathbb{R}^4\times T^2\) from the elliptic genera. \cite{Lee:2020rns} initiated the Cardy limit study of the 6d SCFTs on \(\mathbb{R}^4\times T^2\) for the (2,0) A-type theory and the higher rank E-string theory. In \cite{Lee:2020rns}, the authors computed the 6d free energy by evaluating the elliptic genus summation with the continuum approximation in the Cardy limit. We will mainly follow the same method pioneered in \cite{Lee:2020rns}, but extend their studies to (2,0) D/E-type SCFTs.
 \par 
 The Cardy limit of \(\mathbb{R}^4\times T^2\) is defined as the large momenta limit, i.e., the limit where \(J_{1,2}\) and \(P\) are large in (\ref{eq83_R4T2index}). In terms of the canonical ensemble, this limit can be achieved by setting the conjugate chemical potentials to be small as follows,
 \begin{align}
 |\epsilon_{1,2}| \ll 1, \quad |\beta|\ll 1
 \label{eq496_Cardy}
 \end{align}
where \(\beta\equiv -2\pi i \tau\) is a conjugate chemical potential to \(P\). Now, recall that \(\epsilon_{1,2}\) are the IR regulators on \(\mathbb{R}^4\) by adjusting the effective volume as \(\text{vol}(\mathbb{R}^4)\sim {1\over \epsilon_1 \epsilon_2}\). Also, \(\beta\) is inversely proportional to the radius of the spatial circle in \(T^2\). Therefore, the Cardy limit (\ref{eq496_Cardy}) corresponds to the thermodynamic limit where the spatial volume diverges. As a result, the leading free energy in the Cardy limit should scale as \(\log Z_{\mathbb{R}^4\times T^2}\sim \mathcal{O}({1\over \epsilon_1 \epsilon_2 \beta})\) which is the spatial volume factor of the 6d background. 
\par  
In general, one may consider the complexified chemical potentials for \(\epsilon_{1,2}\) and \(\beta\). Such non-trivial phases of the chemical potentials are important to observe the deconfining behavior of the superconformal indices \cite{Choi:2018vbz}, including 6d SCFTs on \(S^5\times S^1\) \cite{Nahmgoong:2019hko}. However, on \(\mathbb{R}^4\times T^2\), one can observe the deconfining free energy even at the real chemical potential setting. It can be achieved by setting the $\Omega$-background parameters to be close to the self-dual point \(\epsilon_1=-\epsilon_2\) by giving them different signs. Therefore, for simplicity, we shall consider the following chemical potential setting in the rest of this subsection,
\begin{align}
\epsilon_1>0,\quad \epsilon_2<0,\quad \beta>0,\quad  \text{Re}[m]=0.
\label{eq528_ChemSet}
\end{align} 
Note that the flavor chemical potential \(m\) is purely imaginary with \(\text{Im}[m]\) of order \(\mathcal{O}(1)\). Also, we will set the tensor VEVs \(\mathbf{v}\) to be sufficiently close to the origin of the tensor branch \(\mathbf{v}=0\). More precisely, we will assume that \(\mathbf{v}\ll \beta^{-1}\).
\par 
We expect that the leading volume divergence \(\log Z_{\mathbb{R}^4\times T^2}\sim \mathcal{O}({1\over \epsilon_1 \epsilon_2 \beta})\) does not depend on the order of taking the Cardy limit (\ref{eq496_Cardy}). Hence, let us take the small \(\beta\) limit first. In this limit, the asymptotic form of the elliptic genus can be easily computed from the modular property. Using (\ref{eq173_index}) and (\ref{eq178_EG}), the S-dual transformation of the elliptic genus can be written as follows,
\begin{align}
Z_{\mathfrak{n}}(\tau|\epsilon_{1,2},m)&=\exp\Bigr[{1\over \beta}\Bigr({\epsilon_1 \epsilon_2\over 2} \sum_{a,b}\Omega_{a,b}n_a n_b+(m^2-\epsilon_+^2)\sum_a n_a \Bigr) \Bigr]
\cdot 
Z_{\mathfrak{n}}(-{1\over \tau}|{\epsilon_{1,2}\over \tau},{m\over \tau}).
\label{eq535_SDUAL}
\end{align}
Now, we consider the asymptotics of the dual elliptic genus \(Z_{\mathfrak{n}}(-{1\over \tau}|{\epsilon_{1,2}\over \tau},{m\over \tau})\) in the Cardy limit. In the right hand side of (\ref{eq535_SDUAL}), the dual instanton fugacity \(q^D=e^{-4\pi^2/\beta}\) is much smaller than \(1\). Therefore, one might guess that the dual instanton corrections in \(Z_{\mathfrak{n}}(-{1\over \tau}|{\epsilon_{1,2}\over \tau},{m\over \tau})\) can be always ignored in the Cardy limit. However, one should be careful due to the presence of other fugacities such as \(e^{\pm m/\tau}\) which can give additional growth factor that can overcome the dual instanton suppression. After carefully investigating the suppression/growth factors from each fugacities, we found that asymptotics of the dual elliptic genus \(Z_{\mathfrak{n}}(-{1\over \tau}|{\epsilon_{1,2}\over \tau},{m\over \tau})\)is given as follows,
\begin{align}
 Z_{\mathfrak{n}}(-{1\over \tau}|{\epsilon_{1,2}\over \tau},{m\over \tau})
 =\exp\Bigr[{-2\pi i m(2p+1)-4\pi^2 (p+p^2)+\mathcal{O}(\epsilon_{1,2})\over \beta}\sum_a n_a +o(\beta^{-1})\Bigr]
 \label{eq541_dualAsymp}
\end{align}
which we checked for the results given in the appendix \ref{append_EG}. Here, \(p\) is an integer such that \(2\pi p<\text{Im}[m]<2\pi (p+1)\). When \(p=0\), the dual elliptic genus is dominated by the dual perturbative part at \(\mathcal{O}((q^D)^0)\). When \(p\neq 0\), the dominant contribution in the dual elliptic genus comes from the \((p+p^2)\sum n_a\) number of dual instantons. As we shall show later, considering the effect of dual instantons at generic \(p\) is important to recover the \(2\pi i\) periodicity of \(m\).
\par 
For simplicity, let us first consider the \(p=0\) chamber where \(0<\text{Im}[m]<2\pi \). By plugging (\ref{eq541_dualAsymp}) into (\ref{eq535_SDUAL}), one obtains  the following form,
\begin{align}
\log Z_{\mathfrak{n}}={1\over \beta}\Bigr( {\epsilon_1 \epsilon_2\over 2}\sum_{a,b}\Omega_{a,b} n_a n_b +m(m-2\pi i)\sum_a n_a+\mathcal{O}(\epsilon_{1,2})\Bigr)+o(\beta^{-1}).
\end{align}
Now, let us evaluate the 6d \(\mathbb{R}^4\times T^2\) index by summing over the elliptic genera as follows,
\begin{align}
Z_{\mathbb{R}^4\times T^2}&=Z_0\sum_{\mathfrak{n}} Z_{\mathfrak{n}} e^{-\mathfrak{n}\cdot \mathbf{v}}
\nonumber \\
&=Z_0\sum_{\mathfrak{n}}\exp\Bigr[  {1\over \beta}\Bigr( {\epsilon_1 \epsilon_2\over 2}\sum_{a,b}\Omega_{a,b} n_a n_b +m(m-2\pi i)\sum_a n_a+\mathcal{O}(\epsilon_{1,2})\Bigr)+o(\beta^{-1})\Bigr].
\label{eq554_Summand}
\end{align}
Note that the effect of the string fugacity \(e^{-\mathfrak{n}\cdot \mathbf{v}}\) is subleading since we assumed that \(\mathbf{v}\ll \beta^{-1}\). Also, the summation is convergent since \(\text{det}({\epsilon_1 \epsilon_2\over \beta}\Omega)<0\). Now, let us further take the remaining Cardy limit, i.e., \(\epsilon_{1,2}\to 0\) limit. We take a continuum approximation of the string number by defining following continuous variables,
\begin{align}
x_a \equiv -\epsilon_1 \epsilon_2 n_a.
\end{align}
Although \(n_a\) is a discrete variable with \(\Delta n_a=1\), \(x_a\) can be viewed as an almost continuous variable since \(\Delta x_a = -\epsilon_1 \epsilon_2 \ll 1\) in the Cardy limit. Then, the elliptic genus summation (\ref{eq554_Summand}) can be approximated to the integral as follows,
\begin{align}
Z_{\mathbb{R}^4\times T^2}=Z_0 \int_0^\infty [\prod_{a=1}^{r_{\mathfrak{g}}}dx_a]
\exp\Bigr[{1\over \epsilon_1 \epsilon_2 \beta}\Bigr( {1\over 2}\sum_{a,b}\Omega_{a,b}x_a x_b +m(2\pi i-m)\sum_a x_a\Bigr)+o({1\over \epsilon_1 \epsilon_2 \beta}) \Bigr].
\end{align}
The free energy \(\log Z_{\mathbb{R}^4\times T^2}\) can be evaluated from the saddle point approximation of the above integral.  The peak of the Gaussian is located at 
\begin{align}
x_a&=m(m-2\pi i)\sum_b (\Omega^{-1})_{a,b}+\mathcal{O}(\epsilon_{1,2}).
\label{eq559_peak}
\end{align}
Here, notice that the vector \(\sum_b (\Omega^{-1})_{a,b} \mathbf{e}_b = \rho\) is the Weyl vector of the Lie algebra \(\mathfrak{g}\), i.e. \(\rho={1\over 2}\sum_{\alpha\in \Psi^+_{\mathfrak{g}}} \alpha\). For the simply-laced Lie algebras, the Weyl vector is given as follows,
\begin{align}
2\rho
&=\sum_{a=1}^r a(r+1-a)\mathbf{e}_a, &&\mathfrak{g}=A_r
\nonumber \\
&=\sum_{a=1}^{r-2} a(2r-1-a)\mathbf{e}_a+{r(r-1)\over 2}(\mathbf{e}_{r-1}+\mathbf{e}_r), &&\mathfrak{g}=D_r
\nonumber \\
&=(16, 30, 42, 30, 16, 22),& &\mathfrak{g}=E_6
\nonumber \\
&=(34,66,96,75,52,27,49),& &\mathfrak{g}=E_7
\nonumber \\
&=(92, 182, 270, 220, 168, 114, 58, 136),&&\mathfrak{g}=E_8.
\label{eq611_WEYL}
\end{align}
In the Cardy limit, the string number \(\mathfrak{n}\) has a non-zero VEV on the tensor branch, given by \(\langle n_a\rangle=-{x_a\over \epsilon_1 \epsilon_2}\) \cite{Lee:2020rns}. Therefore, the Weyl vector of \(\mathfrak{g}\) determines the distribution of the self-dual strings in 6d (2,0) \(\mathfrak{g}\)-type SCFT in the Cardy limit.
\par 
The contribution of the elliptic genera is localized near the peak (\ref{eq559_peak}), and the leading free energy can be obtained from the value of the integrand at the peak. As a result, we obtain the following free energy, 
\begin{align}
\log Z_{\mathbb{R}^4\times T^2}\simeq \log Z_0 -{\sum_{a,b}(\Omega^{-1})_{a,b} \over 2} {m^2(2\pi i-m)^2\over \epsilon_1 \epsilon_2 \beta}
\end{align}
where we ignored the subleading corrections in the Cardy limit. From the expression of \(Z_0\) given in (\ref{eq280_Z0}), one can compute that \(\log Z_0=-{r_{\mathfrak{g}}\over 24}{m^2(2\pi i -m)^2\over \epsilon_1 \epsilon_2 \beta}\) \cite{Kim:2017zyo}. Also, the group theoretic constant \(\sum_{a,b}(\Omega)^{-1}_{a,b}\) is equal to \({h_{\mathfrak{g}}^\vee d_{\mathfrak{g}}\over 12}\) where \(h_{\mathfrak{g}}^\vee\) is a dual Coxeter number and \(d_{\mathfrak{g}}^\vee\) is a dimension of the Lie algebra \({\mathfrak{g}}\). Therefore, we obtain the Cardy formula of 6d (2,0) \({\mathfrak{g}}\)-type theory on \(\mathbb{R}^4\times T^2\) as follows, 
\begin{align}
\log Z_{\mathbb{R}^4\times T^2}\simeq -{h_{\mathfrak{g}}^\vee d_{\mathfrak{g}} +r_{\mathfrak{g}}\over 24} {m^2(2\pi i-m)^2\over \epsilon_1 \epsilon_2 \beta}.
\end{align}
For the simply-laced Lie algebra, the overall constant \(h_{\mathfrak{g}}^\vee d_{\mathfrak{g}}+r_{\mathfrak{g}}\) is given as follows,
\begin{align}
h_{\mathfrak{g}}^\vee d_{\mathfrak{g}}+r_{\mathfrak{g}}
&=(r+1)^3-1, & &{\mathfrak{g}}=A_r
\nonumber \\
&=4r^3-6r^2+3r, & &{\mathfrak{g}}=D_r
\nonumber \\
&= 942, \ 2401, \ 7448, & &{\mathfrak{g}}=E_{6,7,8}
\end{align}
Note that in our chemical potential setting (\ref{eq528_ChemSet}) and \(0<\text{Im}[m]<2\pi\), we can see that \(\log Z \gg 1\), which signals the macroscopic number of deconfining degrees of freedom. The above result is a straightforward generalization of the (2,0) A-type formula in \cite{Lee:2020rns} to general ADE-type theories. The overall factor of the free energy \(h_{\mathfrak{g}}^\vee d_{\mathfrak{g}}+r_{\mathfrak{g}}\) was also observed in the 6d Cardy formula on \(S^5\times S^1\) \cite{Nahmgoong:2019hko}, which explains the black hole entropy in the dual gravity.
\par 
Lastly, let us consider the free energy in the general chamber of \(m\) given by \(2\pi p < \text{Im}[m]<2\pi (p+1)\). In this chamber, one should use a general form (\ref{eq541_dualAsymp}) with \(p\neq 0\). Inserting (\ref{eq541_dualAsymp}) to (\ref{eq535_SDUAL}) yields,
\begin{align}
\log Z_{\mathfrak{n}}={1\over \beta}\Bigr( {\epsilon_1 \epsilon_2\over 2}\sum_{a,b}\Omega_{a,b} n_a n_b +(m-2\pi i p)(m-2\pi i(p+1) )\sum_a n_a+\mathcal{O}(\epsilon_{1,2})\Bigr)+o(\beta^{-1}).
\end{align}
After following the similar calculations explained so far, we obtain the following 6d free energy,
\begin{align}
\log Z_{\mathbb{R}^4\times T^2}\simeq -{h_G^\vee d_G +r_G\over 24} {(m-2\pi i p)^2(2\pi i(p+1)-m)^2\over \epsilon_1 \epsilon_2 \beta}
\end{align}
One can see that the above expression is periodic under \(m\sim m+2\pi i\). Therefore, as we emphasized, the instanton correction in the dual elliptic genus (\ref{eq541_dualAsymp}) is essential to recover the periodicity of \(m\).

\subsection{(2,0) Superconformal index and \(\mathcal{W}\)-algebra}
In this subsection, we compute the superconformal indices of the 6d (2,0) SCFTs from the elliptic genus method. We will focus on the unrefined limit of the chemical potentials where the superconformal indices are expected to be reduced to the \(\mathcal{W}\)-algebra character \cite{Beem:2014kka}. We show that our results agree with the prediction, including E-type cases.
\par 
Let us briefly review the superconformal index of 6d (2,0) SCFT. See \cite{Kim:2016usy} for a detailed review. The bosonic part of the 6d (2,0) superconformal algebra is given by \(SO(6,2)\) conformal symmetry and \(SO(5)\) R-symmetry. Let us denote \(E\) and \(J_{1,2,3}\) as the charges for \(SO(6,2)\) and \(Q_{1,2}\) as the charges for \(SO(5)\), which are all normalized to be \(\pm {1\over 2}\) for spinors. We take a defining supercharge of the index to give the BPS bound \(E \geq 2Q_1+2Q_2+J_1+J_2+J_3\). Then, the superconformal index is defined as follows,
\begin{align}
I = \text{Tr}\Bigr[(-1)^F e^{-\omega_1 J_1-\omega_2 J_2- \omega_3 J_3}e^{-\Delta_1 Q_1-\Delta_2 Q_2} \Bigr]
\label{eq656_SCI}
\end{align}
where the trace is taken over the Hilbert space of the radially quantized 6d SCFT on \(S^5\times \mathbb{R}_\text{time}\). The chemical potentials are constrained by \(\Delta_1+\Delta_2-\omega_1-\omega_2-\omega_3=0\) to preserve the supersymmetry. We will use the notation \(\Delta_R=\Delta_1+\Delta_2\) and \(\Delta_L=\Delta_1-\Delta_2\) in this subsection. 
\par 
The 6d superconformal index was studied from localization of 5d MSYM on \(S^5\) \cite{Kim:2012ava, Kallen:2012zn,Kim:2012qf,Lockhart:2012vp,Kim:2012tr,Kim:2013nva,Minahan:2013jwa}. The path integral on \(S^5\) is localized at the three fixed points, and the contribution from each fixed point is given by the \(\mathbb{R}^4\times T^2\) index on the tensor branch. The full \(S^5\times S^1\) index can be obtained from the three copies of \(\mathbb{R}^4\times T^2\) indices as follows \cite{Kim:2016usy},
\begin{align}
I =  {e^{-E_0-S_\text{bkgd}}\over W_{\mathfrak{g}}  \sqrt{|\Omega_{\mathfrak{g}} |}}
\int [\prod_{a=1}^r dv_a] e^{-S_0} \hat{Z}_{[1]} \hat{Z}_{[2]} \hat{Z}_{[3]}.
\label{eq617_SCI}
\end{align}
Let us explain the various factors in (\ref{eq617_SCI}). First, \(\hat{Z}_{[k]}\) is the Weyl-symmetrized \(\mathbb{R}^4\times T^2\) index. It is basically same with the \(\mathbb{R}^4\times T^2\) index in (\ref{eq286_INST}) discussed in this paper so far, but multiplied with the Weyl-symmetric factor as follows,
\begin{align}
&\hat{Z}=Z_\text{Weyl} \cdot Z,\quad  Z_\text{Weyl}=\text{PE}\Bigr[{1\over 2}{\sinh{m\pm \epsilon_+\over 2}\over \sinh{\epsilon_{1,2}\over 2} }
\sum_{\alpha\in\Psi^+_{\mathfrak{g}} }(e^{\alpha(\mathbf{v})}-e^{-\alpha(\mathbf{v})} ) \Bigr]. 
\label{eq674_Weyl}
\end{align}
The reason that we introduce the Weyl symmetric factor is that the elliptic genus expansion \(Z=Z_0 \sum_{\mathfrak{n}} Z_{\mathfrak{n}}e^{-\mathfrak{n}\cdot \mathbf{v}}\) is not symmetric under the Weyl reflection \(\mathbf{v}\to -\mathbf{v}\). By multiplying the prefactor \(Z_\text{Weyl}\), one can make the full index \(\hat{Z}=Z_\text{Weyl}Z\) symmetric under the Weyl reflection. Also, the contribution at the $k^{\text{th}}$ fixed point on \(S^5\) is given by \(\hat{Z}_{[k]}=\hat{Z}_{\mathbb{R}^4\times T^2} (\tau_{[k]}, \ v_{[k]},  \ \epsilon_{1,[k]}, \ \epsilon_{2,[k]}, \ m_{[k]})\), and the chemical potentials are given as follows \cite{Kim:2016usy},
\begin{align}
\tau_{[k]}&={2\pi i  \over \omega_k}, &  
v_{[k],a}&=  {2\pi \omega\over \omega_k}v_a,
\nonumber \\
\epsilon_{1,[k]}&=2\pi i({\omega_{k+1}\over \omega_k}-1),&  
\epsilon_{2,[k]}&=2\pi i({\omega_{k-1}\over \omega_k}-1),  &  
m_{[k]}&=-\pi i( {\Delta_L\over \omega_k}+1 ).
\label{eq727_par}
\end{align}
where \(\omega={\omega_1+\omega_2+\omega_3\over 3}\). Second, \(S_0\) is the classical action of the 5d MSYM on the squashed \(S^5\), and it is given as follows, 
\begin{align}
S_0={2\pi^2 \omega^2 \over \omega_1 \omega_2 \omega_3} \sum_{a,b=1}^r (\Omega^{-1})_{a,b} v_a v_b.
\end{align}
Third, \(S_\text{bkgd}\) is the background action which gives a divergent behavior when the chemical potentials are small \(\Delta_{1,2}\sim \omega_{1,2,3}\ll 1\). Its form is given as follows \cite{Kim:2012qf, Chang:2019uag},
\begin{align}
S_\text{bkgd}=r_{\mathfrak{g}}{\pi^2\over 24}    { \omega_1^2+\omega_2^2+\omega_3^2-2\omega_1 \omega_2 -2\omega_2 \omega_3-2\omega_3 \omega_1-\Delta_L^2 \over \omega_1 \omega_2 \omega_3 }.
\label{eq492_background}
\end{align}
Also, \(E_0\) is the Casimir energy which gives a divergent behavior when the chemical potentials are large \(\Delta_{1,2}\sim \omega_{1,2,3}\gg 1\). Its form is given as follows \cite{Bobev:2015kza},
\begin{align}
E_0&={h_G^\vee d_G\over 384} { (\Delta_R^2-\Delta_L^2)^2 \over \omega_1 \omega_2 \omega_3 }
+r_G{\Delta_R+\Delta_L\over 48} 
\nonumber \\
&+ {r_G\over 384} {(\Delta_R-\Delta_L) (2\omega_1-\Delta_R-\Delta_L)(2\omega_2-\Delta_R-\Delta_L)(2\omega_3-\Delta_R-\Delta_L)\over \omega_1 \omega_2 \omega_3 } .
\label{eq513_casimir}
\end{align}
Lastly, \(|\Omega_\mathfrak{g}|\) is the determinant of the Cartan matrix, and \(W_{\mathfrak{g}}\) is the dimension of the Weyl group of \(\mathfrak{g}\). For the simply-laced Lie algebra, they are given as follows,
\begin{align}
\mathfrak{g}&=A_r, & |\Omega_{\mathfrak{g}}|&=r+1, & W_{\mathfrak{g}}=&(r+1)!
\nonumber \\
\mathfrak{g}&=D_r, & |\Omega_{\mathfrak{g}}|&=4, & W_{\mathfrak{g}}=&2^{r-1}r!
\nonumber \\
\mathfrak{g}&=E_6, & |\Omega_{\mathfrak{g}}|&=3, & W_{\mathfrak{g}}=&51,840 
\nonumber \\
\mathfrak{g}&=E_7, & |\Omega_{\mathfrak{g}}|&=2, & W_{\mathfrak{g}}=&2,903,040 
\nonumber \\
\mathfrak{g}&=E_8, & |\Omega_{\mathfrak{g}}|&=1, & W_{\mathfrak{g}}=&696,729,600 .
\end{align}
\par 
Although the 6d superconformal indices have been studied extensively in various places, the proper prescription to evaluate the integral (\ref{eq617_SCI}) is currently unknown when the chemical potentials are fully refined. The main reason is that the chemical potentials in the integrand of (\ref{eq617_SCI}) is `S-dualized' in the sense that \(\tau_{[k]}\sim {1\over \omega_k}\). Therefore, the integral expression (\ref{eq617_SCI}) does not have a proper fugacity expansion structure with respect to \(e^{-\omega_{1,2,3}}\) and \(e^{-\Delta_{1,2}}\).
\par 
Therefore, we will study the index (\ref{eq656_SCI}) with unrefined chemical potentials. Eventually, we would take a single chemical potential to be independent by setting \(\Delta_1=2\omega\), \(\Delta_2=\omega\), and \(\omega_{1,2,3}=\omega\). However, directly substituting these unrefined values makes the integrand of (\ref{eq617_SCI}) singular due to the vanishing \(\Omega\)-background. Therefore, we shall first consider the following unrefinement,
\begin{align}
\Delta_L=\omega_1+\omega_2-\omega_3,
\end{align}
and then take \(\omega_{1,2,3}\to \omega\) limit later. In the above setting, the chemical potentials of each \(\mathbb{R}^4\times T^2\) indices in the integrand satisfy a simple relation. According to (\ref{eq727_par}), we obtain
\begin{align}
e^{-m_{[1]}}=e^{+\epsilon_{-,[1]}},\quad 
e^{-m_{[2]}}=e^{-\epsilon_{-,[2]}},\quad 
e^{-m_{[3]}}=e^{+\epsilon_{+,[3]}}
\end{align}
Here, \(e^{-m}=e^{\pm \epsilon_+}\) and \(e^{-m}=e^{\pm \epsilon_-}\) are special points that the elliptic genera become much simplified. When \(e^{-m}=e^{\pm \epsilon_+}\), all elliptic genera we obtain become \(0\) unless the base degree is empty. When \(e^{-m}=e^{\pm \epsilon_-}\), all elliptic genera we obtain become \(+1\), \(-1\), or \(0\). For all the data we obtained, we check that the following two properties hold,
\begin{align}
&e^{-m}=e^{\pm \epsilon_+}: \ \sum_{\mathfrak{n}}Z_{\mathfrak{n}} e^{-\mathfrak{n}\cdot \mathbf{v}}=1
\nonumber \\
&e^{-m}=e^{\pm \epsilon_-}: \ \sum_{\mathfrak{n}}Z_{\mathfrak{n}}e^{-\mathfrak{n}\cdot \mathbf{v}} 
=\prod_{\alpha\in \Psi^+_{\mathfrak{g}} }(1-e^{-\alpha\cdot \mathbf{v}}).
\label{eq727_SF}
\end{align}
According to (\ref{eq727_SF}), the maximal base degree for the non-vanishing elliptic genus when \(e^{-m}=e^{\pm \epsilon_-}\) is expected to be \(\mathfrak{n}=2\rho\) where \(\rho\) is a Weyl vector (\ref{eq611_WEYL}). Unfortunately, we could not reach that bound due to the limited computing power. However, our data in appendix \ref{append_EG} still provide non-trivial evidence for (\ref{eq727_SF}) inclulding E-type cases.
\par 
Similarly, the pure momentum contribution \(Z_0\) in (\ref{eq280_Z0}) and the Weyl prefactor in (\ref{eq674_Weyl}) become
\begin{align}
&e^{-m}=e^{\pm \epsilon_+}: &&Z_0=q^{r_{\mathfrak{g}}\over 24}\eta(\tau)^{-r_{\mathfrak{g}} } &&Z_\text{Weyl}=1,
\nonumber \\
&e^{-m}=e^{\pm \epsilon_-}: &&Z_0=1 &&Z_\text{Weyl}=\prod_{\alpha \in\Psi^+_G}\Bigr({ 1-e^{\alpha(\mathbf{v})}  \over 1-e^{-\alpha(\mathbf{v})} } \Bigr)^{1\over 2} .
\end{align}
Then, one can rewrite the integral expression (\ref{eq617_SCI}) as follows,
\begin{align}
I&=e^{-{h_{\mathfrak{g}}^\vee d_{\mathfrak{g}}\over 24}{\omega_3(\omega_1+\omega_2)^2 \over \omega_1 \omega_2} }[e^{-{\omega_3 \over 24}}\eta({2\pi i \over \omega_3}  ) ]^{-r_{\mathfrak{g}} }
\nonumber \\
&\times {1 \over W_{\mathfrak{g}} \sqrt{|\Omega_{\mathfrak{g}}|}}\int [\prod_{a=1}^{r_{\mathfrak{g}}} dv_a] e^{-S_0} \prod_{\alpha\in \Psi^+_{\mathfrak{g}}}
\Bigr(2\sinh {\pi \omega \alpha(\mathbf{v})\over \omega_1} \Bigr)
\Bigr(2\sinh {\pi \omega \alpha(\mathbf{v})\over \omega_2} \Bigr).
\label{eq747_GAUS}
\end{align}
Now, the integrand in (\ref{eq747_GAUS}) becomes Gaussian, and the evaluation is straightforward. After taking the further unrefinement \(\omega_{1,2,3}=\omega\) and using the S-duality of the Dedekind eta function \(\eta({2\pi i \over \omega})=\sqrt{\omega\over 2\pi}\eta(-{\omega\over 2\pi i}) \), we obtain that
\begin{align}
I=\text{PE}\Bigr[r_{\mathfrak{g}} {e^{-\omega}\over 1-e^{-\omega}}\Bigr] \prod_{\alpha\in\Psi_{\mathfrak{g}}^+} (1-e^{-\omega \alpha\cdot \rho})
\end{align}
where \(\rho={1\over 2}\sum_{\alpha\in \Psi_{\mathfrak{g}}^+} \alpha\) is a Weyl vector. The above unrefined index can be also written in the following form,
\begin{align}
I=\text{PE}\Bigr[ {\sum_{n\in \mathfrak{D}_{\mathfrak{g}}} e^{-n\omega}\over 1-e^{-\omega}} \Bigr].
\label{eq789_SCI}
\end{align}
Here, \(\mathfrak{D}_{\mathfrak{g}}\) is a set of dimensions of the Casimir operators of the Lie algebra \(\mathfrak{g}\), and it is given as follows,
\begin{align}
&\mathfrak{D}_{A_N}=\{2,3,...,N \}, &&\mathfrak{D}_{D_N}=\{2,4,6,...,2N-2,N \}
\nonumber \\
&\mathfrak{D}_{E_6}=\{2,5,6,8,9,12 \},&&\mathfrak{D}_{E_7}=\{2,6,8,10,12,14,18 \},
\nonumber \\
&\mathfrak{D}_{E_8}=\{2,8,12,14,18,20,24,30 \}.
\end{align}
\par 
The unrefined limit corresponds to the `chiral algebra limit' of the 6d theory, and the superconformal index (\ref{eq789_SCI}) is the same with the \(\mathcal{W}\)-algebra character of type \(\mathfrak{g}\) \cite{Beem:2014kka}. The unrefined superconformal index (\ref{eq789_SCI}) was previously computed for A/D-type (2,0) theories with the instanton partition function method. However, it had been still a conjecture that the index should become the \(\mathcal{W}\)-algebra character for E-type also. In this paper, we partially check that the conjecture is true by computing the self-dual strings' elliptic genera in (2,0) E-type theories.

\section{Concluding remarks}
\label{sec:Conclusion}
In this paper, we studied the elliptic genera of the self-dual strings in general 6d (2,0) SCFTs. We found that the elliptic genus is invariant under the flip symmetry, which is a novel kinematical constraint that we found in this paper. Using the 5d perturbative data as an input, we obtained elliptic genera at low base degrees using the modular property and the flip symmetry. Then, we could further conjecture the vanishing bound of the GV invariants that enables us to bootstrap the elliptic genera up to arbitrary base degrees. 
\par 
We did various consistency checks with our results from the bootstrap. For the elliptic genera of (2,0) A-type theories, our result is exactly the same as the localization computation in \cite{Haghighat:2013}. The general formulas for the elliptic genera in D/E-type theories are currently unknown. However, we check that our results agree with the instanton partition function of 5d maximal SYM with the D-type gauge group \cite{Hwang:2014uwa, Hwang:2016gfw}. For E-type theories, our results give the first non-perturbative computation beyond the 5d perturbative data. 
\par 
As straightforward applications, we utilize our data to compute the 6d Cardy formula on \(\mathbb{R}^4\times T^2\) \cite{Lee:2020rns} and the unrefined (2,0) superconformal index, which is previously computed for A, or D-type theories only. There seem to be many other important topics that our results can be used. Here, we finish our paper by listing some of them.
\par 
First, it would be interesting to construct the instanton partition functions of 5d E-type maximal SYM from the elliptic genera of the 6d (2,0) E-type SCFT. In general, one can study the instanton partition function from the ADHM construction \cite{Atiyah:1978ri} of the instanton quantum mechanics if the gauge group is classical. Unlike classical groups, computing the instantons of the exceptional gauge group has been a difficult problem. Although instanton partition functions in many 5d exceptional gauge theories were computed \cite{DelZotto:2016pvm, Kim:2018gjo, DelZotto:2018tcj, Kim:2019uqw}, there has been no known result for the maximal SYM case. However, the string fugacity expansion of the instanton partition function can be computed from our data in appendix \ref{append_EG}. Therefore, if one can construct an appropriate ansatz for the instanton partition functions of E-type maximal SYM, the coefficients can be fitted by comparing with the elliptic genera.
\par 
Second, one can also consider the extension to 6d little string theories (LSTs), which are non-local QFTs given by the UV-completion of 6d (2,0) SCFTs, or (1,1) SYMs. They also admit affine ADE-type classification, and the elliptic genera were computed in \cite{Kim:2015gha} for A-type LSTs. Especially in \cite{Kim:2018gak}, the authors bootstrapped the elliptic genera of type LSTs using the T-duality between (2,0) and (1,1) LSTs. However, the T-duality bootstrap is incomplete for D/E-type LSTs. It would be interesting if our flip-symmetry or the vanishing condition for the GV invariants can give noan-trivial results for the D/E-type LSTs also.
\par
Third, our result could provide hints to other methods of computing the elliptic genera of (2,0) SCFTs. In particular, it would be nice to see if one could write down 6d blow-up equations that the elliptic genera are supposed to satisfy, along the line of \cite{Gu_2019,Gu_2019II,gu2019elliptic,Gu:2020fem}.
\par 
Lastly, let us make a few comments on the modular bootstrap of the elliptic genera in 6d (1,0) SCFTs. In this paper, we could bootstrap all the elliptic genera in (2,0) SCFTs only from the 5d perturbative data and the modular anomaly. However, a similar bootstrap procedure is known to be impossible for general (1,0) SCFTs with gauge groups, unless we know the precise vanishing condition \cite{DelZotto:2017mee}. Still, one can utilize the flip symmetry to bootstrap the elliptic genera in (1,0) theory if it preserves \(SU(2)\) flavor symmetry originating from the (2,0) \(SO(5)\) R-symmetry. If the flip symmetry and other kinematical constraints can fix the elliptic genera of (1,0) SCFTs at low base degrees, it would be helpful to conjecture the precise vanishing bound from them to initiate the bootstrap program.


\section*{Acknowledgements}
We are grateful to Kihong Lee, Kimyeong Lee, Amir-Kian Kashani-Poor, Joonho Kim, Seok Kim, and Xin Wang for helpful discussions. We would like to thank Amir-Kian Kashani-Poor, Seok Kim, and Kimyeong Lee for their reading of the draft and providing inspiring comments. In particular, we thank Amir-Kian Kashani-Poor for kindly improving English writing of the draft. JN is supported by KIAS Individual Grant PG076401. ZD is supported by KIAS Individual Grant PG076901.

\newpage 
\appendix

\section{Modular Forms}\label{App:MF}
    In this appendix, we collect some standard facts about the theory of modular forms \cite{Zagierbook, serre2012course, EZ}.
\subsection*{Elliptic Modular Forms}
\begin{mydef}
Suppose $k$ is an integer. A function $f: \mathbb{H} \rightarrow \mathbb{C}$ is called a modular form of weight $k$ for full modular group $\text{SL}(2, \mathbb{Z})$ if f is holomorphic on $\mathbb{H} \cup \{\infty\}$ and satisfies the following equation
\begin{equation}
    \label{modular}
f(\frac{a \tau + b}{c \tau + d}) = (c\tau + d)^{k} f(\tau),\ \mathrm{for\ any} \begin{pmatrix} a & b\\
c & d\\
\end{pmatrix}\in \text{SL}(2, \mathbb{Z}) 
\end{equation}
\end{mydef}

In particular, if we choose the matrix $T = \begin{pmatrix} 1 & 1\\
0 & 1\\
\end{pmatrix}$, we find that $f(\tau + 1) = f(\tau)$. Thus if we introduce $q = \exp (2\pi i \tau)$, $f$ can be expanded as a power series at $\tau = i\infty$,
\begin{equation}
f(\tau) = \sum\limits_{n \geq 0} a_{n} q^{n}\,.
\end{equation}
The condition $n \geq 0$ is guaranteed by the holomorphicity at the infinity. If furthermore $a_{n} = 0$, then $f$ is called a $\mathit{cusp}$ form.

A natural construction of modular forms involves summing over all two dimensional lattice points, which are reshuffled under the $\text{SL}(2, \mathbb{Z})$ action. Indeed, we can define normalized Eisenstein series,
\begin{equation}
    E_{k}(\tau) = \frac{1}{\zeta(k)}\sum_{\substack{(m, n)\in \mathbb{Z}^2\\(m,n) \neq (0,0)}} \frac{1}{(m + n\tau)^k}\,,
\end{equation}
and it is easy to show that $E_k$ is a modular form of weight $k$ when $k$ is an integer larger than two.\footnote{This is needed to ensure that we can rearrange the order of summation.} When $k$ is odd, it trivially vanishes so we only need to focus on $k$ being even. The overall constant is chosen such that when  power expanded in terms of $q$, the leading constant of $E_k$ is one. In fact, we have,
\begin{equation}
    E_k = 1 - \frac{2k}{B_k}\sum\limits_{n = 1}^{\infty}\sigma_{k - 1}(n)\, q^n,
\end{equation}
where $B_k$ is the $k^{\text{th}}$ Bernoulli number and $\sigma_{i}(n)$ denotes the sum of the $i^\text{th}$ powers of the positive divisors of $n$.

It is a classical result that the ring of holomorphic (elliptic) modular forms $M_\ast = \oplus_{k \geq 0} M_{k}(\text{SL}(2, \mathbb{Z}))$ is freely generated by $E_4$ and $E_6$ \cite{Zagierbook, serre2012course}.

Functions that are not exactly modular are also important. In the main text, we also need the Dedekind $\eta$ function,
\begin{equation}
\eta(\tau) = q^\frac{1}{24}\prod\limits_{i = 1}^{\infty}(1 - q^i)\,.
\end{equation}
Its $24^{\text{th}}$ power is a cusp modular form of weight twelve. It can be expressed in terms of the two generators,
\begin{equation}
\eta(\tau)^{24} = \frac{1}{1728}(E_4(\tau)^3 - E_6(\tau)^2) = q - 24q^2 + \cdots\,.
\end{equation}
Another important function that is nearly modular is the second Eisenstein series $E_2$. Under $\text{SL}(2, \mathbb{Z})$, it transforms as
\begin{equation}\label{App:E2}
    E_2(\frac{a \tau + b}{c \tau + d})=(c \tau +d)^2 E_2(\tau) -\frac{6 i}{\pi} c(c\tau+d)\, .
\end{equation}
$E_2$ is a famous example of quasimodular forms. In fact, we can enlarge our ring $M_\ast$ to be the ring of quasi-modular forms $\tilde{M}_\ast$, by relaxing the condition (\ref{modular}) to have lower powers in $(c\tau + d)$ on the right-hand side. $\tilde{M}_\ast$ is also finitely generated. Actually, it can be shown \cite{Zagierbook} that $\tilde{M}_\ast$ is generated from $M_\ast$ just by adding one extra generator $E_2$.

If one wants to maintain modularity, we can add an extra piece to $E_2$ and define
\begin{equation}
    \hat E_2(\tau)=E_2(\tau)-\frac{3}{\pi {\rm Im}(\tau)}\, .
\end{equation}
$\hat E_2(\tau)$ can be shown to be modular with weight two, but clearly at the expense of losing holomorphicity. It is an example of almost holomorphic modular forms.

\subsection*{Jacobi Modular Forms}
\label{appendixJMF}
\begin{mydef}\label{defjmf}
A Jacobi modular form of weight $k$ and index $m$ is a function $\phi:\mathbb{H}\times \mathbb{C}\rightarrow \mathbb{C}$ that depends on a modular 
	parameter $\tau\in \mathbb{H}$ and an elliptic parameter $z\in \mathbb{C}$. It transforms 
	under the action of $\text{SL}(2,\mathbb{Z})$ on $\mathbb{H} \times \mathbb{C}$ as
    \begin{equation}
	\tau \mapsto \tau_\gamma=\frac{a \tau+ b}{c \tau + d}, \quad z \mapsto z_\gamma=\frac{z}{c \tau + d}\quad  {\rm with} \quad  
	\left(\begin{array}{cc}a&b\\ c& d \end{array}\right) \in {\rm SL}(2, \mathbb{Z})  \ ,
	\label{PSL2} 
    \end{equation} 
	as  
	\begin{align}\label{modtran} 
    \phi_{k,m}\left(\tau_\gamma, z_\gamma\right) &= (c \tau + d)^k e^{\frac{2 \pi i m c z^2}{c \tau + d}} \phi_{k,m}(\tau,z)\,, \\
	\phi_{k,m}(\tau,z +\lambda \tau+ \mu) &= e^{- 2 \pi i m (\lambda^2 \tau+ 2 \lambda z)}\phi_{k,m}(\tau,z) \quad \forall  \lambda, \mu \in \mathbb{Z}\,,\label{elltrans}
	\end{align}
\end{mydef}
(\ref{modtran}) is known as the \emph{modular} transformation which is a generalization of modular transform, while the second one (\ref{elltrans}) is known as the \emph{elliptic} transform.

From the definition, if we choose $T = \begin{pmatrix} 1 & 1\\
0 & 1\\
\end{pmatrix}$ and $\lambda = 0, \mu = 1$ in equations (\ref{modtran}) and (\ref{elltrans}) respectively, we see that the Jacobi form is invariant under the shift $\tau \rightarrow \tau + 1$ and $z\rightarrow z + 1$, hence it enjoys a double Fourier expansion,
\begin{equation} 
	\phi(\tau,z)=\sum_{n,r} c(n,r) q^n y^r, \qquad {\rm for} \ \ q=e^{2 \pi i \tau},\ \  y=e^{2 \pi i z} \ .
\end{equation}

It can be shown that $c(n,r)$ only depends on $r$ and an $\text{SL}(2, \mathbb{Z})$ invariant combination $4nm - r^2$. In other words, $c(n,r)=C(4 n m-r^2,r)$. We can further define three classes of Jacobi modular forms: holomorphic Jacobi forms $J^{h}_{\ast,\ast}$ satisfy the constraint $c(n,r) = 0$ unless $4nm \geq r^2$, cusp forms $J^{c}_{\ast,\ast}$ satisfy $c(n,r) = 0$ unless $4nm > r^2$ and weak Jacobi forms $J^{w}_{\ast,\ast}$ satisfy $c(n,r) = 0$ unless $n \geq 0$. Clearly we have,
\begin{equation}
    J^{c}_{\ast,\ast} \subset J^{h}_{\ast,\ast} \subset J^{w}_{\ast,\ast}.
\end{equation}

An important theorem in \cite{EZ} shows that weak Jacobi forms $J^{w}_{\ast,\ast}$ of integer index is freely generated over the ring of elliptic modular forms by two generators $\varphi_{-2,1} (\tau,z)$ and $\varphi_{0,1} (\tau,z)$. 

$\varphi_{-2,1}$ and $\varphi_{0,1}$ can be defined in terms of the Jacobi theta functions. For $a$ and $b$ $\in \{0, 1/2\}$, we have
\begin{equation}
    \Theta\left[a \atop b\right](\tau,z)=\sum_{n\in \mathbb{Z}} e^{\pi i (n + a)^2 \tau + 2 \pi i z (n+a) + 2 \pi i bn}\ . 
\end{equation}

The four theta functions in our convention are
\begin{equation}
    \theta_1=i\Theta\left[\tfrac{1}{2} \atop \tfrac{1}{2}\right], \quad \theta_2=\Theta\left[\tfrac{1}{2} \atop 0\right], \quad \theta_3=\Theta\left[0 \atop 0\right], \quad \theta_4=\Theta\left[0 \atop \frac{1}{2}  \right]\,.
\end{equation}

Based on $\theta_i(\tau, z)$, we can give explicit forms of our two generators,
\begin{equation}
\begin{aligned}
    \varphi_{-2,1}(\tau,z) &= \frac{ \theta_1(\tau,z)^2} {\eta(\tau)^6} \, , \\ 
	\varphi_{0,1}(\tau,z) &= 4\Big(\frac{ \theta_2(\tau,z)^2} {  \theta_2(\tau, 0)^2 } +\frac{ \theta_3(\tau,z)^2} {\theta_3(\tau, 0)^2 } +\frac{ \theta_4(\tau,z)^2} {  \theta_4(\tau, 0)^2 } \Big). 
\end{aligned}
\end{equation}

We also give the first few terms in their $q$-expansion for convenience,
\begin{equation}
\begin{aligned}
 \phi_{-2,1}(\tau,z) &= -(y - 2 + 1/y) + 2(y^2 -4y +6 -4/y + 1/y^2)\,q + \cdots\,,\\
 \phi_{0,1}(\tau,z) &= (y + 10 + 1/y) + (10y^2 - 64y + 108 - 64/y + 10/y^2)\,q + \dots\,.
\end{aligned}
\end{equation}

Finally, we point out an interesting relation between Jacobi forms $\phi(\tau, z)$ and quasimodular forms $E_2$. Looking back to the modular transformation \eqref{modtran}, the exponential factor on the right-hand side looks a bit annoying, but actually one can introduce an automorphic correction to cancel it. Recall the modular transformation of $E_2$ \eqref{App:E2}, it is easy to verify that the combination $\Phi(\tau,z) = e^{\frac{m\pi^2}{3}E_2 z^2}\phi(\tau, z)$ transforms without that factor. In other words, if we consider a series expansion of $\Phi(\tau,z)$ in $z$,
\begin{equation}
    \Phi(\tau,z) = \sum_{n \geq 0} g_n(\tau) z^n\,,
\end{equation}
Each $g_n(\tau)$ now behaves nicely as genuine elliptic modular forms. From this we can extract a differential equation for $\phi(\tau, z)$,
\begin{equation}\label{App:DEJF}
    \frac{\partial}{\partial E_2} \Phi(\tau,z) = 0 \quad \Rightarrow \quad \left(\frac{\partial}{\partial E_2} + \frac{m \pi^2}{3} z^2\right) \phi(\tau, z) = 0\,.
\end{equation}

\section{Elliptic genera}
\label{append_EG}
In this section, we summarize the elliptic genera of the ADE M-strings of a few base degrees.  We consider base degrees with \(n_i\neq 0\) for all \(1\leq i \leq r_{\mathfrak{g}}\) so that the full chain does not degenerate into the shorter pieces of the M-string chain. We used a shorthand notation such that
\begin{align}
\varphi_{-n,1}({\epsilon_+\over 2\pi i})=R_n,\quad 
\varphi_{-n,1}({\epsilon_-\over 2\pi i})=L_n,\quad 
\varphi_{-n,1}({m\over 2\pi i})=f_n.
\end{align}
Also, we further use the identity that
\begin{align}
{1\over 12}\Bigr( \varphi_{-2,1}({x\over 2\pi i})\varphi_{0,1}({y\over 2\pi i})- \varphi_{-2,1}({y\over 2\pi i})\varphi_{0,1}({x\over 2\pi i})\Bigr)
= \theta(x\pm y)
\end{align}
where we defined a function \(\theta(x)\) as follows,
\begin{align}
\theta(x)\equiv {\theta_1(\tau,{x\over 2\pi i} )\over \eta(\tau)^3}
\end{align}
and use a notation that \(\theta(x\pm y)=\theta(x+y)\theta(x-y)\). Finally, in the case when the numerator of a higher rank elliptic genus is factorized into that of a lower rank elliptic genus, we write it as \((\text{higher rank base degree})=\text{factor}\times (\text{lower rank base degree})\) in the tables. We find that attaching a single M-string to another single M-string yields \(\theta(m\pm \epsilon_-)\) factor universally.

\begin{table}[ht!]
\centering
\begin{tabular}{|c||c|}\hline
\(\mathfrak{n}\) & \(\mathcal{N}\) \\ \hline\hline
\( \begin{pmatrix}
1
\end{pmatrix}\)
&
\( \theta(m\pm \epsilon_+)  \)
\\ \hline
\( \begin{pmatrix}
2
\end{pmatrix}\)
&  
{\tiny 
\(
{1\over 2^{11}3^6} \theta(m\pm \epsilon_+) \theta(m\pm \epsilon_-)
( 27 R_2^6 E_4{}^3-45 R_2^4 R_0^2 E_4{}^2-24 R_2^5 R_0 E_4 E_6-15 R_2^2 \
R_0^4 E_4-32 R_2^6 E_6{}^2-40 R_2^3 R_0^3 E_6+R_0^6) 
\)
}
\\
\hline  
\end{tabular}
\caption{Elliptic genera of \(A_{1,2,3}\). The result that cannot be obtained from the flip symmetry is marked with the asterisk.}
\label{tabEGA1}
\end{table}

\begin{table}[ht!]
\centering
\begin{tabular}{|c||c|}\hline
\(\mathfrak{n}\) & \(\mathcal{N}\) \\ \hline\hline
\(\begin{pmatrix}
&1&\\
1&1&1
\end{pmatrix} \)
&
\(\theta(m\pm\epsilon_+)\cdot \theta(m\pm\epsilon_-)^3\)
\\ \hline
\(\begin{pmatrix}
&1&\\
2&1&1
\end{pmatrix}  \)
&  
\(
\theta(m\pm\epsilon_-)^2\times 
\begin{pmatrix}
2 & 1
\end{pmatrix}
\)
\\
\hline  
\end{tabular}
\caption{Elliptic genera of \(D_4\). The result that cannot be obtained from the flip symmetry is marked with the asterisk. }
\label{tabEGD4}
\end{table}

\begin{table}[ht!]
\centering
\begin{tabular}{|c||c|}\hline
\(\mathfrak{n}\) & \(\mathcal{N}\) \\ \hline\hline
\(\begin{pmatrix}
&&1&\\
1&1&1&1
\end{pmatrix} \)
&
\( \theta(m\pm \epsilon_+)\cdot \theta(m\pm \epsilon_-)^4 \)
\\ \hline
\(\begin{matrix}
\begin{pmatrix}
&&1&\\
1&1&1&2
\end{pmatrix}  
 \\
\begin{pmatrix}
&&1&\\
2&1&1&1
\end{pmatrix}  
\end{matrix}\)
&
\(
\theta(m\pm \epsilon_-)^3 \times 
\begin{pmatrix}
2 & 1
\end{pmatrix}
\)
\\ \hline 
\(\begin{pmatrix}
&&1&\\
1&2&1&1
\end{pmatrix}  \)
&\(\theta(m\pm \epsilon_-)^2\times  
\begin{pmatrix}
1&2&1
\end{pmatrix}
\)  
\\ \hline
\(\begin{pmatrix}
&&1&\\
1&1&2&1
\end{pmatrix} \)
&
\(\theta(m\pm \epsilon_-)\times  
\begin{pmatrix}
&1& \\
1&2&1
\end{pmatrix}
\)
\\
\hline 
\end{tabular}
\caption{Elliptic genera of \(D_5\). The result that cannot be obtained from the flip symmetry is marked with the asterisk.}
\label{tabEGD5}
\end{table}

\begin{table}[ht!]
\centering
\begin{minipage}[t!]{.49\columnwidth}
\begin{tabular}{|c||c|}\hline
\(\mathfrak{n}\) & \(\mathcal{N}\) \\ \hline\hline
\(\begin{matrix}
\begin{pmatrix}
&&1&& \\
1 & 1 & 1 & 1 & 1
\end{pmatrix}
\end{matrix}\)
& \(\theta(m\pm \epsilon_+)\cdot \theta(m\pm \epsilon_-)^5 \) 
\\ \hline
\(
\begin{matrix}
\begin{pmatrix}
&&2&& \\
1&1&1&1&1
\end{pmatrix} \\
\begin{pmatrix}
&&1&& \\
2&1&1&1&1
\end{pmatrix}
\end{matrix}
\)
&
\(
\theta(m\pm \epsilon_-)^4 \times
\begin{pmatrix}
2 & 1
\end{pmatrix}
\)  
\\ \hline
\(
\begin{matrix}
\begin{pmatrix}
&&1&& \\
1&2&1&1&1
\end{pmatrix}
\end{matrix}
\)
&
\(
\theta(m\pm \epsilon_-)^3 \times
\begin{pmatrix}
1 & 2 & 1
\end{pmatrix}
\)   
\\ \hline
\(
\begin{matrix}
\begin{pmatrix}
&&1&& \\
1&1&2&1&1
\end{pmatrix}
\end{matrix}
\)
&
\(
\theta(m\pm \epsilon_-)^2 \times
\begin{pmatrix}
&1& \\
1 & 2 & 1
\end{pmatrix}
\)
\\ \hline 
\end{tabular}
\caption{Elliptic genera of \(E_6\)}
\label{tabEGE6}
\end{minipage}
\begin{minipage}[t!]{.49\columnwidth}
\begin{tabular}{|c||c|}\hline
\(\mathfrak{n}\) & \(\mathcal{N}\) \\ \hline\hline
\(\begin{pmatrix}
&&1&&& \\
1&1 & 1 & 1 & 1 & 1
\end{pmatrix}\)
& \(\theta(m\pm \epsilon_+)\cdot \theta(m\pm \epsilon_-)^6\) 
\\ \hline
\(
\begin{matrix}
\begin{pmatrix}
&&2&&& \\
1&1&1&1&1&1
\end{pmatrix} \\
\begin{pmatrix}
&&1&&& \\
2&1&1&1&1&1
\end{pmatrix} \\
\begin{pmatrix}
&&1&&& \\
1&1&1&1&1&2
\end{pmatrix}
\end{matrix}
\)
&
\(
\theta(m\pm \epsilon_-)^5 \times
\begin{pmatrix}
2 & 1
\end{pmatrix}
\)   
\\ \hline
\(
\begin{matrix}
\begin{pmatrix}
&&1&&& \\
1&2&1&1&1&1
\end{pmatrix}\\
\begin{pmatrix}
&&1&&& \\
1&1&1&2&1&1
\end{pmatrix} \\
\begin{pmatrix}
&&1&&& \\
1&1&1&1&2&1
\end{pmatrix}
\end{matrix}
\)
&
\(
\theta(m\pm \epsilon_-)^4 \times
\begin{pmatrix}
1 & 2 & 1
\end{pmatrix}
\)  
\\ \hline
\(
\begin{matrix}
\begin{pmatrix}
&&1&&& \\
1&1&2&1&1&1
\end{pmatrix}
\end{matrix}
\)
&
\(
\theta(m\pm \epsilon_-)^3 \times
\begin{pmatrix}
&1& \\
1 & 2 & 1
\end{pmatrix}
\)
\\ \hline 
\end{tabular}
\caption{Elliptic genera of \(E_7\)}
\label{tabEGE7}
\end{minipage}
\end{table}

\begin{table}[ht!]
\centering
\begin{tabular}{|c||c|}\hline
\(\mathfrak{n}\) & \(\mathcal{N}\) \\ \hline\hline
\(\begin{pmatrix}
&&1&&&& \\
1&1&1 & 1 & 1 & 1 & 1
\end{pmatrix}\)
& \(\theta(m\pm \epsilon_+)\cdot \theta(m\pm \epsilon_-)^7\) 
\\ \hline
\(
\begin{matrix}
\begin{pmatrix}
&&2&&&& \\
1&1&1&1&1&1&1
\end{pmatrix} \\
\begin{pmatrix}
&&1&&&& \\
2&1&1&1&1&1&1
\end{pmatrix} \\
\begin{pmatrix}
&&1&&&& \\
1&1&1&1&1&1&2
\end{pmatrix}
\end{matrix}
\)
&
\(
\theta(m\pm \epsilon_-)^6 \times
\begin{pmatrix}
2 & 1
\end{pmatrix}
\)  
\\ \hline
\(
\begin{matrix}
\begin{pmatrix}
&&1&&&& \\
1&2&1&1&1&1&1
\end{pmatrix}\\
\begin{pmatrix}
&&1&&&& \\
1&1&1&2&1&1&1
\end{pmatrix}  \\
\begin{pmatrix}
&&1&&&& \\
1&1&1&1&2&1&1
\end{pmatrix}    \\
\begin{pmatrix}
&&1&&&& \\
1&1&1&1&1&2&1
\end{pmatrix}  
\end{matrix}
\)
&
\(
\theta(m\pm \epsilon_-)^5 \times
\begin{pmatrix}
1 & 2 & 1
\end{pmatrix} 
\) 
\\ \hline
\(
\begin{matrix}
\begin{pmatrix}
&&1&&&& \\
1&1&2&1&1&1&1
\end{pmatrix}
\end{matrix}
\)
&
\(
\theta(m\pm \epsilon_-)^4 \times
\begin{pmatrix}
&1& \\
1 & 2 & 1
\end{pmatrix}
\)
\\ \hline 
\end{tabular}
\caption{Elliptic genera of \(E_8\)}
\label{tabEGE8}
\end{table}

 \clearpage 
 
\section{GV invariants}\label{appen:BPS}
Here, we list some GV invariants of geometries that engineer E-type (2,0) SCFTs, extracted from the elliptic genera.

\begin{table}[ht!]
\centering
\begin{tabular}{|c|cccc|}\hline
      $\{(1,1,2,1,1,1),e,2\}$ & $e=$1&2&3&4 \\ \hline
     $g$ = 0& -158& -14198 & -630198 & -18294632 \\
         1& 34 & 8323 & 605826 & 24543838\\
         2& & -1506& -224344 & -13830246\\
         3& & 84 & 40334& 4281922\\
         4& & & -3640 & -801172 \\
         5& & & 140 & 92994\\
         \hline
\end{tabular}

\begin{tabular}{|c|cccc|}\hline
      $\{(1,1,2,1,1,1),e,3\}$ & $e=$1&2&3&4 \\ \hline
     $g$ = 0& 51 & 6090 & 311415 & 9876000\\
         1& -8 & -3098 & -274390 & -12459206\\
         2& & 477& 91798 & 6540187\\
         3& & -24 & -14764& -1868740\\
         4& & & 1190 & 319875 \\
         5& & & -40 & -33634\\
         \hline
\end{tabular}
\begin{tabular}{|c|cccc|}\hline
      $\{(1,1,2,1,1,1),e,7\}$ & $e=$1&2&3&4 \\ \hline
     $g$ = 0& & -38 & -5266 & -296958\\
         1& & 5 & 2498 & 252881\\
         2& & & -370 & -82390\\
         3& & & 18& 13096\\
         4& & &  & -1046 \\
         5& & &  & 34\\ 
         \hline
\end{tabular}
\caption{Unrefined GV invariants from (2,0) \(E_6\) SCFT at base degree (1,1,2,1,1,1), fiber degree 1 to 4 and mass degree 2, 3 and 7.}

\end{table}

\begin{table}[ht!]
\centering
\begin{tabular}{|c|ccccc|}\hline
    $\{(1,1,2,1,1,1),1,1\}$ & $2j_+=$0&1&2&3&4\\ \hline
     $2j_-=$0 & 55 && 28 && 1\\
     1&& 33 && 4 &\\
     \hline
\end{tabular}

\begin{tabular}{|c|cccccc|}\hline
    $\{(1,1,2,1,1,1),2,2\}$ & $2j_+=$0&1&2&3&4&5\\ \hline
     $2j_-=$0 &&1305&& 315&& 6\\
     1&900&& 740&& 71&\\
     2&& 301&& 94 && 4 \\
     3&6 &&16 && 6 &\\
     \hline
\end{tabular}

\caption{Refined GV invariants from (2,0) \(E_6\) SCFT at base degree (1,1,2,1,1,1), fiber degree 1, 2 with mass degree 1 and 2 respectively.}
\end{table}

\begin{table}[ht!]
\centering
\begin{tabular}{|c|cccc|}\hline
      $\{(1,1,2,1,1,1,1),e,1\}$ & $e=$1&2&3&4 \\ \hline
     $g$ = 0& 323& 27832 & 1319194 & 42422226 \\
         1& -86 & -17200 & -1289146 & -56879466\\
         2& & 3244& 482499 & 31860566\\
         3& & -168 & -86166& -9715442\\
         4& & & 7490 & 1768586 \\
         5& & & -280 & -197964\\
         \hline
\end{tabular}

\begin{tabular}{|c|cccc|}\hline
      $\{(1,1,2,1,1,1,1),e,4\}$ & $e=$1&2&3&4 \\ \hline
     $g$ = 0& -10 & -2216 & -166600 & -7098368\\
         1& 1 & 868 & 124584 & 7922736\\
         2& & -94& -33822 & -3580000\\
         3& & 3 & 4175& 851158\\
         4& & & -240 & -116466 \\
         5& & & 5 & 9321\\
         \hline
\end{tabular}

\begin{tabular}{|c|cccc|}\hline
      $\{(1,1,2,1,1,1,1),e,5\}$ & $e=$1&2&3&4 \\ \hline
     $g$ = 0& 1 & 438 & 45561 & 2346531\\
         1& & -118 & -28154 & -2317034\\
         2& & 7 & 5970 & 902530\\
         3& & & -528& -178930\\
         4& & & 17&  19447 \\
         5& & &  & -1122\\ 
         \hline
\end{tabular}
\caption{Unrefined GV invariants from (2,0) \(E_7\) SCFT at base degree (1,1,2,1,1,1,1), fiber degree 1 to 4 and mass degree 1, 4 and 5.}
\end{table}

\begin{table}[ht!]
    \centering
 \begin{tabular}{|c|ccccc|}\hline
         $\{(1,1,2,1,1,1,1),1,1\}$&$2j_+=$0&1&2&3&4 \\\hline
         $2j_-=$0&59&&29&&1\\
         1&& 35&&4&\\
         \hline
    \end{tabular}    

 \begin{tabular}{|c|cccc|}\hline
         $\{(1,1,2,1,1,1,1),2,4\}$&$2j_+=$0&1&2&3  \\\hline
         $2j_-=$0&& 346&& 54\\
         1&201&& 111&\\
         2&&30&&4\\
         3&0&&1&\\
         \hline
    \end{tabular}     
    \caption{Refined GV invariants from (2,0) \(E_7\) SCFT at base degree (1,1,2,1,1,1,1), fiber degree 1 and 2 with mass degree 1 and 4 respectively.}
\end{table}

\begin{table}[ht!]
\centering
\begin{tabular}{|c|cccc|}\hline
      $\{(1,1,1,1,1,1,1,1),e,3\}$ & $e=$2&3&4&5 \\ \hline
     $g$ = 0& 1100 & 57254 & 1851595 & 44039385\\
         1& -378 & -37522 & -1771932 & -55325096\\
         2& 21& 7462& 611859 & 27249080\\
         3& & -406 & -88228& -6517634\\
         4& & & 4256 & 745339 \\
         5& & & & -31626\\
         \hline
\end{tabular}

\begin{tabular}{|c|cccc|}\hline
      $\{(1,1,1,1,1,1,1,1),e,4\}$ & $e=$2&3&4&5 \\ \hline
     $g$ = 0& -224 & -18256 & -731712 & -19082144\\
         1& 42 & 9604 & 617120 & 22839016\\
         2& & -1344& -179592 & -10100384\\
         3& & 35 & 20146& 2097858\\
         4& & & -630 & -197036 \\
         5& & & & 6146\\
         \hline
\end{tabular}
\begin{tabular}{|c|cccc|}\hline
      $\{(1,1,1,1,1,1,1,1),e,5\}$ & $e=$2&3&4&5 \\ \hline
     $g$ = 0& 21& 3948 & 216538 & 6995758\\
         1& & -1470 & -153062 & -7192724\\
         2& & 105 & 34650 & 2743685\\
         3& & & -2590& -466144\\
         4& & & 35 & 32480 \\
         5& & &  & -602\\ 
         \hline
\end{tabular}
\caption{Unrefined GV invariants from (2,0) \(E_8\) SCFT at base degree (1,1,1,1,1,1,1,1), fiber degree 2 to 5 and mass degree 3, 4 and 5.}
\end{table}

\begin{table}[ht!]
    \centering
    \begin{tabular}{|c|cc|}\hline
         $\{(1,1,1,1,1,1,1,1),0,0\}$&$2j_+=$0&1  \\\hline
         $2j_-=$0&&1\\
         \hline
    \end{tabular}
    
 \begin{tabular}{|c|ccc|}\hline
         $\{(1,1,1,1,1,1,1,1),1,1\}$&$2j_+=$0&1&2  \\\hline
         $2j_-=$0&29&&8\\
         1&& 15&\\
         \hline
    \end{tabular}    

 \begin{tabular}{|c|cccc|}\hline
         $\{(1,1,1,1,1,1,1,1),2,2\}$&$2j_+=$0&1&2&3  \\\hline
         $2j_-=$0&& 443&& 8\\
         1&351&& 169&\\
         2&&77&&0\\
         \hline
    \end{tabular}     
    \caption{Refined GV invariants from (2,0) \(E_8\) SCFT at base degree (1,1,1,1,1,1,1,1), fiber degree 0, 1 and 2 with mass degree 0, 1 and 2 respectively.}
    \label{tab:my_label}
\end{table}

\clearpage

\bibliographystyle{JHEP}
\bibliography{main}

\providecommand{\href}[2]{#2}\begingroup\raggedright\begin{thebibliography}{10}

\bibitem{Gaiotto_2012}
D.~Gaiotto, \emph{N = 2 dualities},
  \href{https://doi.org/10.1007/jhep08(2012)034}{\emph{Journal of High Energy
  Physics} {\bfseries 2012} (2012) }.

\bibitem{Alday:2009aq}
L.~F. Alday, D.~Gaiotto and Y.~Tachikawa, \emph{{Liouville Correlation
  Functions from Four-dimensional Gauge Theories}},
  \href{https://doi.org/10.1007/s11005-010-0369-5}{\emph{Lett. Math. Phys.}
  {\bfseries 91} (2010) 167} [\href{https://arxiv.org/abs/0906.3219}{{\ttfamily
  0906.3219}}].

\bibitem{Dimofte_2013}
T.~Dimofte, D.~Gaiotto and S.~Gukov, \emph{Gauge theories labelled by
  three-manifolds},
  \href{https://doi.org/10.1007/s00220-013-1863-2}{\emph{Communications in
  Mathematical Physics} {\bfseries 325} (2013) 367–419}.

\bibitem{Gadde_2016}
A.~Gadde, S.~Gukov and P.~Putrov, \emph{Fivebranes and 4-manifolds},
  \href{https://doi.org/10.1007/978-3-319-43648-7_7}{\emph{Progress in
  Mathematics} (2016) 155–245}.

\bibitem{Witten:1995zh}
E.~Witten, \emph{{Some comments on string dynamics}},  in \emph{{Future
  perspectives in string theory. Proceedings, Conference, Strings'95, Los
  Angeles, USA, March 13-18, 1995}}, pp.~501--523, 1995,
  \href{https://arxiv.org/abs/hep-th/9507121}{{\ttfamily hep-th/9507121}}.

\bibitem{Seiberg:1996vs}
N.~Seiberg and E.~Witten, \emph{{Comments on string dynamics in
  six-dimensions}},
  \href{https://doi.org/10.1016/0550-3213(96)00189-7}{\emph{Nucl. Phys.}
  {\bfseries B471} (1996) 121}
  [\href{https://arxiv.org/abs/hep-th/9603003}{{\ttfamily hep-th/9603003}}].

\bibitem{Vafa_1996}
C.~Vafa, \emph{Evidence for f-theory},
  \href{https://doi.org/10.1016/0550-3213(96)00172-1}{\emph{Nuclear Physics B}
  {\bfseries 469} (1996) 403–415}.

\bibitem{Morrison_1996}
D.~R. Morrison and C.~Vafa, \emph{Compactifications of f-theory on calabi-yau
  threefolds. (i)},
  \href{https://doi.org/10.1016/0550-3213(96)00242-8}{\emph{Nuclear Physics B}
  {\bfseries 473} (1996) 74–92}.

\bibitem{Morrison_1996II}
D.~R. Morrison and C.~Vafa, \emph{Compactifications of f-theory on calabi-yau
  threefolds (ii)},
  \href{https://doi.org/10.1016/0550-3213(96)00369-0}{\emph{Nuclear Physics B}
  {\bfseries 476} (1996) 437–469}.

\bibitem{Heckman_2014}
J.~J. Heckman, D.~R. Morrison and C.~Vafa, \emph{On the classification of 6d
  scfts and generalized ade orbifolds},
  \href{https://doi.org/10.1007/jhep05(2014)028}{\emph{Journal of High Energy
  Physics} {\bfseries 2014} (2014) }.

\bibitem{Heckman:2015bfa}
J.~J. Heckman, D.~R. Morrison, T.~Rudelius and C.~Vafa, \emph{{Atomic
  Classification of 6D SCFTs}},
  \href{https://doi.org/10.1002/prop.201500024}{\emph{Fortsch. Phys.}
  {\bfseries 63} (2015) 468}
  [\href{https://arxiv.org/abs/1502.05405}{{\ttfamily 1502.05405}}].

\bibitem{Katz:1996fh}
S.~H. Katz, A.~Klemm and C.~Vafa, \emph{{Geometric engineering of quantum field
  theories}}, \href{https://doi.org/10.1016/S0550-3213(97)00282-4}{\emph{Nucl.
  Phys.} {\bfseries B497} (1997) 173}
  [\href{https://arxiv.org/abs/hep-th/9609239}{{\ttfamily hep-th/9609239}}].

\bibitem{Nekrasov:2002qd}
N.~A. Nekrasov, \emph{{Seiberg-Witten prepotential from instanton counting}},
  \href{https://doi.org/10.4310/ATMP.2003.v7.n5.a4}{\emph{Adv. Theor. Math.
  Phys.} {\bfseries 7} (2003) 831}
  [\href{https://arxiv.org/abs/hep-th/0206161}{{\ttfamily hep-th/0206161}}].

\bibitem{Nekrasov}
N.~A. Nekrasov and A.~Okounkov, \emph{Seiberg-witten theory and random
  partitions}, \href{https://doi.org/10.1007/0-8176-4467-9_15}{\emph{The Unity
  of Mathematics} 525–596}.

\bibitem{Hollowood:2003cv}
T.~J. Hollowood, A.~Iqbal and C.~Vafa, \emph{{Matrix models, geometric
  engineering and elliptic genera}},
  \href{https://doi.org/10.1088/1126-6708/2008/03/069}{\emph{JHEP} {\bfseries
  03} (2008) 069} [\href{https://arxiv.org/abs/hep-th/0310272}{{\ttfamily
  hep-th/0310272}}].

\bibitem{IKV}
A.~Iqbal, C.~Kozcaz and C.~Vafa, \emph{{The Refined topological vertex}},
  \href{https://doi.org/10.1088/1126-6708/2009/10/069}{\emph{JHEP} {\bfseries
  10} (2009) 069} [\href{https://arxiv.org/abs/hep-th/0701156}{{\ttfamily
  hep-th/0701156}}].

\bibitem{Aganagic:2011mi}
M.~Aganagic, M.~C. Cheng, R.~Dijkgraaf, D.~Krefl and C.~Vafa, \emph{{Quantum
  Geometry of Refined Topological Strings}},
  \href{https://doi.org/10.1007/JHEP11(2012)019}{\emph{JHEP} {\bfseries 11}
  (2012) 019} [\href{https://arxiv.org/abs/1105.0630}{{\ttfamily 1105.0630}}].

\bibitem{Huang:2011qx}
M.-x. Huang, A.-K. Kashani-Poor and A.~Klemm, \emph{{The $\Omega$ deformed
  B-model for rigid $\mathcal{N}=2$ theories}},
  \href{https://doi.org/10.1007/s00023-012-0192-x}{\emph{Annales Henri
  Poincare} {\bfseries 14} (2013) 425}
  [\href{https://arxiv.org/abs/1109.5728}{{\ttfamily 1109.5728}}].

\bibitem{Haghighat:2013}
B.~Haghighat, A.~Iqbal, C.~Kozçaz, G.~Lockhart and C.~Vafa,
  \emph{{M-Strings}},
  \href{https://doi.org/10.1007/s00220-014-2139-1}{\emph{Commun. Math. Phys.}
  {\bfseries 334} (2015) 779}
  [\href{https://arxiv.org/abs/1305.6322}{{\ttfamily 1305.6322}}].

\bibitem{Haghighat:2014pva}
B.~Haghighat, G.~Lockhart and C.~Vafa, \emph{{Fusing E-strings to heterotic
  strings: E+E→H}},
  \href{https://doi.org/10.1103/PhysRevD.90.126012}{\emph{Phys. Rev.}
  {\bfseries D90} (2014) 126012}
  [\href{https://arxiv.org/abs/1406.0850}{{\ttfamily 1406.0850}}].

\bibitem{Cai:2014vka}
W.~Cai, M.-x. Huang and K.~Sun, \emph{{On the Elliptic Genus of Three E-strings
  and Heterotic Strings}},
  \href{https://doi.org/10.1007/JHEP01(2015)079}{\emph{JHEP} {\bfseries 01}
  (2015) 079} [\href{https://arxiv.org/abs/1411.2801}{{\ttfamily 1411.2801}}].

\bibitem{Benini_2013}
F.~Benini, R.~Eager, K.~Hori and Y.~Tachikawa, \emph{Elliptic genera of
  two-dimensional n = 2 gauge theories with rank-one gauge groups},
  \href{https://doi.org/10.1007/s11005-013-0673-y}{\emph{Letters in
  Mathematical Physics} {\bfseries 104} (2013) 465–493}.

\bibitem{Benini:2013xpa}
F.~Benini, R.~Eager, K.~Hori and Y.~Tachikawa, \emph{{Elliptic Genera of 2d N =
  2 Gauge Theories}},
  \href{https://doi.org/10.1007/s00220-014-2210-y}{\emph{Commun. Math. Phys.}
  {\bfseries 333} (2015) 1241}
  [\href{https://arxiv.org/abs/1308.4896}{{\ttfamily 1308.4896}}].

\bibitem{Haghighat:2013tka}
B.~Haghighat, C.~Kozcaz, G.~Lockhart and C.~Vafa, \emph{{Orbifolds of
  M-strings}}, \href{https://doi.org/10.1103/PhysRevD.89.046003}{\emph{Phys.
  Rev.} {\bfseries D89} (2014) 046003}
  [\href{https://arxiv.org/abs/1310.1185}{{\ttfamily 1310.1185}}].

\bibitem{HKLV:2014}
B.~Haghighat, A.~Klemm, G.~Lockhart and C.~Vafa, \emph{{Strings of Minimal 6d
  SCFTs}}, \href{https://doi.org/10.1002/prop.201500014}{\emph{Fortsch. Phys.}
  {\bfseries 63} (2015) 294} [\href{https://arxiv.org/abs/1412.3152}{{\ttfamily
  1412.3152}}].

\bibitem{Kim:2014dza}
J.~Kim, S.~Kim, K.~Lee, J.~Park and C.~Vafa, \emph{{Elliptic Genus of
  E-strings}}, \href{https://doi.org/10.1007/JHEP09(2017)098}{\emph{JHEP}
  {\bfseries 09} (2017) 098} [\href{https://arxiv.org/abs/1411.2324}{{\ttfamily
  1411.2324}}].

\bibitem{Gadde:2015tra}
A.~Gadde, B.~Haghighat, J.~Kim, S.~Kim, G.~Lockhart and C.~Vafa, \emph{{6d
  String Chains}}, \href{https://doi.org/10.1007/JHEP02(2018)143}{\emph{JHEP}
  {\bfseries 02} (2018) 143}
  [\href{https://arxiv.org/abs/1504.04614}{{\ttfamily 1504.04614}}].

\bibitem{Kim:2015fxa}
J.~Kim, S.~Kim and K.~Lee, \emph{{Higgsing towards E-strings}},
  \href{https://arxiv.org/abs/1510.03128}{{\ttfamily 1510.03128}}.

\bibitem{Kim:2016foj}
H.-C. Kim, S.~Kim and J.~Park, \emph{{6d strings from new chiral gauge
  theories}},  \href{https://arxiv.org/abs/1608.03919}{{\ttfamily 1608.03919}}.

\bibitem{Kim:2018gjo}
H.-C. Kim, J.~Kim, S.~Kim, K.-H. Lee and J.~Park, \emph{{6d strings and
  exceptional instantons}},  \href{https://arxiv.org/abs/1801.03579}{{\ttfamily
  1801.03579}}.

\bibitem{Kim:2015jba}
S.-S. Kim, M.~Taki and F.~Yagi, \emph{{Tao Probing the End of the World}},
  \href{https://doi.org/10.1093/ptep/ptv108}{\emph{PTEP} {\bfseries 2015}
  (2015) 083B02} [\href{https://arxiv.org/abs/1504.03672}{{\ttfamily
  1504.03672}}].

\bibitem{Hayashi:2017jze}
H.~Hayashi and K.~Ohmori, \emph{{5d/6d DE instantons from trivalent gluing of
  web diagrams}}, \href{https://doi.org/10.1007/JHEP06(2017)078}{\emph{JHEP}
  {\bfseries 06} (2017) 078}
  [\href{https://arxiv.org/abs/1702.07263}{{\ttfamily 1702.07263}}].

\bibitem{Foda_2018}
O.~Foda and R.-D. Zhu, \emph{An elliptic topological vertex},
  \href{https://doi.org/10.1088/1751-8121/aae654}{\emph{Journal of Physics A:
  Mathematical and Theoretical} {\bfseries 51} (2018) 465401}.

\bibitem{Huang:2013yta}
M.-X. Huang, A.~Klemm and M.~Poretschkin, \emph{{Refined stable pair invariants
  for E-, M- and $[p, q]$-strings}},
  \href{https://doi.org/10.1007/JHEP11(2013)112}{\emph{JHEP} {\bfseries 11}
  (2013) 112} [\href{https://arxiv.org/abs/1308.0619}{{\ttfamily 1308.0619}}].

\bibitem{Haghighat:2014vxa}
B.~Haghighat, A.~Klemm, G.~Lockhart and C.~Vafa, \emph{{Strings of Minimal 6d
  SCFTs}}, \href{https://doi.org/10.1002/prop.201500014}{\emph{Fortsch. Phys.}
  {\bfseries 63} (2015) 294} [\href{https://arxiv.org/abs/1412.3152}{{\ttfamily
  1412.3152}}].

\bibitem{Gu_2019}
J.~Gu, B.~Haghighat, K.~Sun and X.~Wang, \emph{Blowup equations for 6d scfts.
  part i}, \href{https://doi.org/10.1007/jhep03(2019)002}{\emph{Journal of High
  Energy Physics} {\bfseries 2019} (2019) }.

\bibitem{Gu_2019II}
J.~Gu, A.~Klemm, K.~Sun and X.~Wang, \emph{Elliptic blowup equations for 6d
  scfts. part ii. exceptional cases},
  \href{https://doi.org/10.1007/jhep12(2019)039}{\emph{Journal of High Energy
  Physics} {\bfseries 2019} (2019) }.

\bibitem{gu2019elliptic}
J.~Gu, B.~Haghighat, A.~Klemm, K.~Sun and X.~Wang, \emph{Elliptic blowup
  equations for 6d scfts. iii: E-strings, m-strings and chains},  2019.

\bibitem{Gu:2020fem}
J.~Gu, B.~Haghighat, A.~Klemm, K.~Sun and X.~Wang, \emph{{Elliptic Blowup
  Equations for 6d SCFTs. IV: Matters}},
  \href{https://arxiv.org/abs/2006.03030}{{\ttfamily 2006.03030}}.

\bibitem{Morrison:2012np}
D.~R. Morrison and W.~Taylor, \emph{{Classifying bases for 6D F-theory
  models}}, \href{https://doi.org/10.2478/s11534-012-0065-4}{\emph{Central Eur.
  J. Phys.} {\bfseries 10} (2012) 1072}
  [\href{https://arxiv.org/abs/1201.1943}{{\ttfamily 1201.1943}}].

\bibitem{DelZotto:2016pvm}
M.~Del~Zotto and G.~Lockhart, \emph{{On Exceptional Instanton Strings}},
  \href{https://doi.org/10.1007/JHEP09(2017)081}{\emph{JHEP} {\bfseries 09}
  (2017) 081} [\href{https://arxiv.org/abs/1609.00310}{{\ttfamily
  1609.00310}}].

\bibitem{Gu:2017ccq}
J.~Gu, M.-x. Huang, A.-K. Kashani-Poor and A.~Klemm, \emph{{Refined BPS
  invariants of 6d SCFTs from anomalies and modularity}},
  \href{https://doi.org/10.1007/JHEP05(2017)130}{\emph{JHEP} {\bfseries 05}
  (2017) 130} [\href{https://arxiv.org/abs/1701.00764}{{\ttfamily
  1701.00764}}].

\bibitem{DelZotto:2017mee}
M.~Del~Zotto, J.~Gu, M.-X. Huang, A.-K. Kashani-Poor, A.~Klemm and G.~Lockhart,
  \emph{{Topological Strings on Singular Elliptic Calabi-Yau 3-folds and
  Minimal 6d SCFTs}},
  \href{https://doi.org/10.1007/JHEP03(2018)156}{\emph{JHEP} {\bfseries 03}
  (2018) 156} [\href{https://arxiv.org/abs/1712.07017}{{\ttfamily
  1712.07017}}].

\bibitem{Kim:2018gak}
J.~Kim, K.~Lee and J.~Park, \emph{{On elliptic genera of 6d string theories}},
  \href{https://doi.org/10.1007/JHEP10(2018)100}{\emph{JHEP} {\bfseries 10}
  (2018) 100} [\href{https://arxiv.org/abs/1801.01631}{{\ttfamily
  1801.01631}}].

\bibitem{DelZotto:2018tcj}
M.~Del~Zotto and G.~Lockhart, \emph{{Universal Features of BPS Strings in
  Six-dimensional SCFTs}},
  \href{https://doi.org/10.1007/JHEP08(2018)173}{\emph{JHEP} {\bfseries 08}
  (2018) 173} [\href{https://arxiv.org/abs/1804.09694}{{\ttfamily
  1804.09694}}].

\bibitem{Duan:2018sqe}
Z.~Duan, J.~Gu and A.-K. Kashani-Poor, \emph{{Computing the elliptic genus of
  higher rank E-strings from genus 0 GW invariants}},
  \href{https://doi.org/10.1007/JHEP03(2019)078}{\emph{JHEP} {\bfseries 03}
  (2019) 078} [\href{https://arxiv.org/abs/1810.01280}{{\ttfamily
  1810.01280}}].

\bibitem{Haghighat:2017vch}
B.~Haghighat, W.~Yan and S.-T. Yau, \emph{{ADE String Chains and Mirror
  Symmetry}}, \href{https://doi.org/10.1007/JHEP01(2018)043}{\emph{JHEP}
  {\bfseries 01} (2018) 043}
  [\href{https://arxiv.org/abs/1705.05199}{{\ttfamily 1705.05199}}].

\bibitem{Shimizu:2016lbw}
H.~Shimizu and Y.~Tachikawa, \emph{{Anomaly of strings of 6d $
  \mathcal{N}=\left(1,0\right) $ theories}},
  \href{https://doi.org/10.1007/JHEP11(2016)165}{\emph{JHEP} {\bfseries 11}
  (2016) 165} [\href{https://arxiv.org/abs/1608.05894}{{\ttfamily
  1608.05894}}].

\bibitem{Bobev_2015}
N.~Bobev, M.~Bullimore and H.-C. Kim, \emph{Supersymmetric casimir energy and
  the anomaly polynomial},
  \href{https://doi.org/10.1007/jhep09(2015)142}{\emph{Journal of High Energy
  Physics} {\bfseries 2015} (2015) }.

\bibitem{Gopakumar:1998ii}
R.~Gopakumar and C.~Vafa, \emph{{M theory and topological strings. 1.}},
  \href{https://arxiv.org/abs/hep-th/9809187}{{\ttfamily hep-th/9809187}}.

\bibitem{Gopakumar:1998jq}
R.~Gopakumar and C.~Vafa, \emph{{M theory and topological strings. 2.}},
  \href{https://arxiv.org/abs/hep-th/9812127}{{\ttfamily hep-th/9812127}}.

\bibitem{Gopakumar_1999}
R.~Gopakumar and C.~Vafa, \emph{On the gauge theory/geometry correspondence},
  \href{https://doi.org/10.4310/atmp.1999.v3.n5.a5}{\emph{Advances in
  Theoretical and Mathematical Physics} {\bfseries 3} (1999) 1415–1443}.

\bibitem{Huang:2013}
M.-X. Huang, A.~Klemm and M.~Poretschkin, \emph{{Refined stable pair invariants
  for E-, M- and $[p, q]$-strings}},
  \href{https://doi.org/10.1007/JHEP11(2013)112}{\emph{JHEP} {\bfseries 11}
  (2013) 112} [\href{https://arxiv.org/abs/1308.0619}{{\ttfamily 1308.0619}}].

\bibitem{Klemm_1997}
A.~Klemm, P.~Mayr and C.~Vafa, \emph{Bps states of exceptional non-critical
  strings}, \href{https://doi.org/10.1016/s0920-5632(97)00422-2}{\emph{Nuclear
  Physics B - Proceedings Supplements} {\bfseries 58} (1997) 177–194}.

\bibitem{BCOVI}
M.~Bershadsky, S.~Cecotti, H.~Ooguri and C.~Vafa, \emph{{Holomorphic anomalies
  in topological field theories}},
  \href{https://doi.org/10.1016/0550-3213(93)90548-4}{\emph{Nucl. Phys.}
  {\bfseries B405} (1993) 279}
  [\href{https://arxiv.org/abs/hep-th/9302103}{{\ttfamily hep-th/9302103}}].

\bibitem{BCOV}
M.~Bershadsky, S.~Cecotti, H.~Ooguri and C.~Vafa, \emph{{Kodaira-Spencer theory
  of gravity and exact results for quantum string amplitudes}},
  \href{https://doi.org/10.1007/BF02099774}{\emph{Commun. Math. Phys.}
  {\bfseries 165} (1994) 311}
  [\href{https://arxiv.org/abs/hep-th/9309140}{{\ttfamily hep-th/9309140}}].

\bibitem{Aganagic_2007}
M.~Aganagic, V.~Bouchard and A.~Klemm, \emph{Topological strings and (almost)
  modular forms},
  \href{https://doi.org/10.1007/s00220-007-0383-3}{\emph{Communications in
  Mathematical Physics} {\bfseries 277} (2007) 771–819}.

\bibitem{HKK}
M.-x. Huang, S.~Katz and A.~Klemm, \emph{{Topological String on elliptic CY
  3-folds and the ring of Jacobi forms}},
  \href{https://doi.org/10.1007/JHEP10(2015)125}{\emph{JHEP} {\bfseries 10}
  (2015) 125} [\href{https://arxiv.org/abs/1501.04891}{{\ttfamily
  1501.04891}}].

\bibitem{Haghighat:2015ega}
B.~Haghighat, S.~Murthy, C.~Vafa and S.~Vandoren, \emph{{F-Theory, Spinning
  Black Holes and Multi-string Branches}},
  \href{https://doi.org/10.1007/JHEP01(2016)009}{\emph{JHEP} {\bfseries 01}
  (2016) 009} [\href{https://arxiv.org/abs/1509.00455}{{\ttfamily
  1509.00455}}].

\bibitem{Kim:2011mv}
H.-C. Kim, S.~Kim, E.~Koh, K.~Lee and S.~Lee, \emph{{On instantons as
  Kaluza-Klein modes of M5-branes}},
  \href{https://doi.org/10.1007/JHEP12(2011)031}{\emph{JHEP} {\bfseries 12}
  (2011) 031} [\href{https://arxiv.org/abs/1110.2175}{{\ttfamily 1110.2175}}].

\bibitem{Bullimore:2014awa}
M.~Bullimore, H.-C. Kim and P.~Koroteev, \emph{{Defects and Quantum
  Seiberg-Witten Geometry}},
  \href{https://doi.org/10.1007/JHEP05(2015)095}{\emph{JHEP} {\bfseries 05}
  (2015) 095} [\href{https://arxiv.org/abs/1412.6081}{{\ttfamily 1412.6081}}].

\bibitem{Hwang:2014uwa}
C.~Hwang, J.~Kim, S.~Kim and J.~Park, \emph{{General instanton counting and 5d
  SCFT}}, \href{https://doi.org/10.1007/JHEP07(2015)063,
  10.1007/JHEP04(2016)094}{\emph{JHEP} {\bfseries 07} (2015) 063}
  [\href{https://arxiv.org/abs/1406.6793}{{\ttfamily 1406.6793}}].

\bibitem{Hwang:2016gfw}
Y.~Hwang, J.~Kim and S.~Kim, \emph{{M5-branes, orientifolds, and S-duality}},
  \href{https://doi.org/10.1007/JHEP12(2016)148}{\emph{JHEP} {\bfseries 12}
  (2016) 148} [\href{https://arxiv.org/abs/1607.08557}{{\ttfamily
  1607.08557}}].

\bibitem{Choi:2018hmj}
S.~Choi, J.~Kim, S.~Kim and J.~Nahmgoong, \emph{{Large AdS black holes from
  QFT}},  \href{https://arxiv.org/abs/1810.12067}{{\ttfamily 1810.12067}}.

\bibitem{Kim:2019yrz}
J.~Kim, S.~Kim and J.~Song, \emph{{A 4d N=1 Cardy Formula}},
  \href{https://arxiv.org/abs/1904.03455}{{\ttfamily 1904.03455}}.

\bibitem{Nahmgoong:2019hko}
J.~Nahmgoong, \emph{{6d superconformal Cardy formulas}},
  \href{https://arxiv.org/abs/1907.12582}{{\ttfamily 1907.12582}}.

\bibitem{KKV}
S.~H. Katz, A.~Klemm and C.~Vafa, \emph{{M theory, topological strings and
  spinning black holes}}, {\emph{Adv. Theor. Math. Phys.} {\bfseries 3} (1999)
  1445} [\href{https://arxiv.org/abs/hep-th/9910181}{{\ttfamily
  hep-th/9910181}}].

\bibitem{PT1}
R.~Pandharipande and R.~P. Thomas, \emph{{Curve counting via stable pairs in
  the derived category}},
  \href{https://doi.org/10.1007/s00222-009-0203-9}{\emph{Invent. Math.}
  {\bfseries 178} (2009) 407}
  [\href{https://arxiv.org/abs/0707.2348}{{\ttfamily 0707.2348}}].

\bibitem{Katz_1996}
S.~Katz, \emph{Gromov-witten invariants via algebraic geometry},
  \href{https://doi.org/10.1016/0920-5632(96)00012-6}{\emph{Nuclear Physics B -
  Proceedings Supplements} {\bfseries 46} (1996) 108–115}.

\bibitem{Choi:2012jz}
J.~Choi, S.~Katz and A.~Klemm, \emph{{The refined BPS index from stable pair
  invariants}}, \href{https://doi.org/10.1007/s00220-014-1978-0}{\emph{Commun.
  Math. Phys.} {\bfseries 328} (2014) 903}
  [\href{https://arxiv.org/abs/1210.4403}{{\ttfamily 1210.4403}}].

\bibitem{Lee:2020rns}
K.~Lee and J.~Nahmgoong, \emph{{Cardy Limits of 6d Superconformal Theories}},
  \href{https://arxiv.org/abs/2006.10294}{{\ttfamily 2006.10294}}.

\bibitem{Choi:2018vbz}
S.~Choi, J.~Kim, S.~Kim and J.~Nahmgoong, \emph{{Comments on deconfinement in
  AdS/CFT}},  \href{https://arxiv.org/abs/1811.08646}{{\ttfamily 1811.08646}}.

\bibitem{Kim:2017zyo}
S.~Kim and J.~Nahmgoong, \emph{{Asymptotic M5-brane entropy from S-duality}},
  \href{https://doi.org/10.1007/JHEP12(2017)120}{\emph{JHEP} {\bfseries 12}
  (2017) 120} [\href{https://arxiv.org/abs/1702.04058}{{\ttfamily
  1702.04058}}].

\bibitem{Beem:2014kka}
C.~Beem, L.~Rastelli and B.~C. van Rees, \emph{{$ \mathcal{W} $ symmetry in six
  dimensions}}, \href{https://doi.org/10.1007/JHEP05(2015)017}{\emph{JHEP}
  {\bfseries 05} (2015) 017} [\href{https://arxiv.org/abs/1404.1079}{{\ttfamily
  1404.1079}}].

\bibitem{Kim:2016usy}
S.~Kim and K.~Lee, \emph{{Indices for 6 dimensional superconformal field
  theories}}, \href{https://doi.org/10.1088/1751-8121/aa5cbf}{\emph{J. Phys.}
  {\bfseries A50} (2017) 443017}
  [\href{https://arxiv.org/abs/1608.02969}{{\ttfamily 1608.02969}}].

\bibitem{Kim:2012ava}
H.-C. Kim and S.~Kim, \emph{{M5-branes from gauge theories on the 5-sphere}},
  \href{https://doi.org/10.1007/JHEP05(2013)144}{\emph{JHEP} {\bfseries 05}
  (2013) 144} [\href{https://arxiv.org/abs/1206.6339}{{\ttfamily 1206.6339}}].

\bibitem{Kallen:2012zn}
J.~Kallen, J.~A. Minahan, A.~Nedelin and M.~Zabzine, \emph{{$N^3$-behavior from
  5D Yang-Mills theory}},
  \href{https://doi.org/10.1007/JHEP10(2012)184}{\emph{JHEP} {\bfseries 10}
  (2012) 184} [\href{https://arxiv.org/abs/1207.3763}{{\ttfamily 1207.3763}}].

\bibitem{Kim:2012qf}
H.-C. Kim, J.~Kim and S.~Kim, \emph{{Instantons on the 5-sphere and
  M5-branes}},  \href{https://arxiv.org/abs/1211.0144}{{\ttfamily 1211.0144}}.

\bibitem{Lockhart:2012vp}
G.~Lockhart and C.~Vafa, \emph{{Superconformal Partition Functions and
  Non-perturbative Topological Strings}},
  \href{https://doi.org/10.1007/JHEP10(2018)051}{\emph{JHEP} {\bfseries 10}
  (2018) 051} [\href{https://arxiv.org/abs/1210.5909}{{\ttfamily 1210.5909}}].

\bibitem{Kim:2012tr}
H.-C. Kim and K.~Lee, \emph{{Supersymmetric M5 Brane Theories on R x CP2}},
  \href{https://doi.org/10.1007/JHEP07(2013)072}{\emph{JHEP} {\bfseries 07}
  (2013) 072} [\href{https://arxiv.org/abs/1210.0853}{{\ttfamily 1210.0853}}].

\bibitem{Kim:2013nva}
H.-C. Kim, S.~Kim, S.-S. Kim and K.~Lee, \emph{{The general M5-brane
  superconformal index}},  \href{https://arxiv.org/abs/1307.7660}{{\ttfamily
  1307.7660}}.

\bibitem{Minahan:2013jwa}
J.~A. Minahan, A.~Nedelin and M.~Zabzine, \emph{{5D super Yang-Mills theory and
  the correspondence to AdS$_7$/CFT$_6$}},
  \href{https://doi.org/10.1088/1751-8113/46/35/355401}{\emph{J. Phys.}
  {\bfseries A46} (2013) 355401}
  [\href{https://arxiv.org/abs/1304.1016}{{\ttfamily 1304.1016}}].

\bibitem{Chang:2019uag}
C.-M. Chang, M.~Fluder, Y.-H. Lin and Y.~Wang, \emph{{Proving the 6d Cardy
  Formula and Matching Global Gravitational Anomalies}},
  \href{https://arxiv.org/abs/1910.10151}{{\ttfamily 1910.10151}}.

\bibitem{Bobev:2015kza}
N.~Bobev, M.~Bullimore and H.-C. Kim, \emph{{Supersymmetric Casimir Energy and
  the Anomaly Polynomial}},
  \href{https://doi.org/10.1007/JHEP09(2015)142}{\emph{JHEP} {\bfseries 09}
  (2015) 142} [\href{https://arxiv.org/abs/1507.08553}{{\ttfamily
  1507.08553}}].

\bibitem{Atiyah:1978ri}
M.~F. Atiyah, N.~J. Hitchin, V.~G. Drinfeld and {\relax Yu}.~I. Manin,
  \emph{{Construction of Instantons}},
  \href{https://doi.org/10.1016/0375-9601(78)90141-X}{\emph{Phys. Lett.}
  {\bfseries A65} (1978) 185}.

\bibitem{Kim:2019uqw}
J.~Kim, S.-S. Kim, K.-H. Lee, K.~Lee and J.~Song, \emph{{Instantons from
  Blow-up}}, \href{https://doi.org/10.1007/JHEP11(2019)092}{\emph{JHEP}
  {\bfseries 11} (2019) 092}
  [\href{https://arxiv.org/abs/1908.11276}{{\ttfamily 1908.11276}}].

\bibitem{Kim:2015gha}
J.~Kim, S.~Kim and K.~Lee, \emph{{Little strings and T-duality}},
  \href{https://doi.org/10.1007/JHEP02(2016)170}{\emph{JHEP} {\bfseries 02}
  (2016) 170} [\href{https://arxiv.org/abs/1503.07277}{{\ttfamily
  1503.07277}}].

\bibitem{Zagierbook}
J.~H. Bruinier, G.~van~der Geer, G.~Harder and D.~Zagier, \emph{The 1-2-3 of
  modular forms}, Universitext. Springer-Verlag, Berlin, 2008,
  \href{https://doi.org/10.1007/978-3-540-74119-0}{10.1007/978-3-540-74119-0}.

\bibitem{serre2012course}
J.-P. Serre, \emph{A course in arithmetic}, vol.~7. Springer Science \&
  Business Media, 2012.

\bibitem{EZ}
M.~Eichler and D.~Zagier, \emph{The theory of {J}acobi forms}, vol.~55 of
  \emph{Progress in Mathematics}. Birkh\"auser Boston, Inc., Boston, MA, 1985,
  \href{https://doi.org/10.1007/978-1-4684-9162-3}{10.1007/978-1-4684-9162-3}.

\end{thebibliography}\endgroup

\end{document}